\documentclass[a4paper,12pt]{article}
\usepackage{authblk}
\usepackage[dvipsnames]{xcolor}
\usepackage{latexsym,amsmath,amssymb,amsfonts,amsthm,dsfont,enumerate,natbib,bbm,threeparttable,rotating, pdflscape,verbatim,amssymb, setspace, graphicx}
\usepackage{algorithm,algcompatible,algpseudocode,booktabs,xcolor,tikz}
\usetikzlibrary{arrows,shapes,chains}
\usepackage[normalem]{ulem}

\usepackage[colorlinks,linkcolor=red,citecolor=blue,urlcolor=blue,breaklinks]{hyperref}
\usepackage{cleveref}

\usepackage{multirow,multicol}

\algnewcommand\algorithmicpara{\textbf{estimate in parallel for}}
\algblockdefx[PARA]{Para}{EndPara}[1]{\algorithmicpara\ #1}{\algorithmicend}
\algnewcommand\algorithmicdivide{\textbf{divide}}
\algnewcommand\algorithmicextract{\textbf{matrix decomposition}}
\algblockdefx[DIVIDE]{Divide}{EndDivide}{\algorithmicdivide}{\algorithmicend}
\algblockdefx[EXTRACT]{Extraction}{EndExtract}{\algorithmicextract}{\algorithmicend}

\algnewcommand{\algorithmicendif}{\textbf{end}}
\algblockdefx[IF]{If}{EndIf}[1]{\algorithmicif\ #1\ \algorithmicthen}{\algorithmicendif}

\newtheorem{theorem}{Theorem}
\newtheorem{lemma}{Lemma}
\newtheorem{condition}{Condition}

\def\var{\text{Var}}

\def\bB{{\bf B}}

\def\bx{{\bf x}}

\newcommand{\bu}{\mbox{\bf u}}

\def\bz{{\bf z}}

\def\0{{\bf 0}}

\newcommand{\btheta}{\mbox{\boldmath $\theta$}}

\newcommand{\bPi}{\mbox{\boldmath $\Pi$}}
\newcommand{\bzero}{\mbox{\bf 0}}
\newcommand{\bSig}{\mbox{\boldmath $\Sigma$}}

\DeclareMathOperator*{\argmax}{arg\,max}
\DeclareMathOperator*{\argmin}{arg\,min}

\RequirePackage{lineno}

\providecommand{\keywords}[1]{\textbf{\textit{Keywords---}} #1}

\begin{document}

\title{Nonparametric Screening for Additive Quantile Regression in Ultra-high Dimension}
\author[1]{Daoji Li\thanks{Corresponding author. Email: dali@fullerton.edu}}
\author[1]{Yinfei Kong}
\author[1]{Dawit Zerom}

\affil[1]{{College of Business and Economics, California State University, Fullerton, CA, 92831, United States}}

\date{ }
\maketitle

\begin{abstract}
In practical applications, one often does not know the “true” structure of the underlying conditional quantile function, especially in the ultra-high dimensional setting. To deal with ultra-high dimensionality, quantile-adaptive marginal nonparametric screening methods have been recently developed.  However, these approaches may miss important covariates that are marginally independent of the response, or may select unimportant covariates due to their high correlations with important covariates. To mitigate such shortcomings, we develop a conditional nonparametric quantile screening procedure (complemented by subsequent selection) for nonparametric additive quantile regression models. Under some mild conditions, we show that the proposed screening method can identify all relevant covariates in a small number of steps with probability approaching one. The subsequent narrowed best subset (via a modified Bayesian information criterion) also contains all the relevant covariates with overwhelming probability. The advantages of our proposed procedure are demonstrated through simulation studies and a real data example..
\end{abstract}


\newpage

\section{Introduction}\label{sec:Intro}

High dimensional data are now frequently collected and analyzed in a large variety of research areas such as genomics, functional magnetic resonance imaging and tomography. For example, one may wish to analyze certain clinical prognosis (such as survival time) using gene expression data where the number of variables (or features) is often much larger than the number of subjects or the sample size.  In this paper, we are interested in identifying variables that truly contribute to the conditional quantiles of the response in the ultra-high dimensional setting.
Existing approaches that rely only on conditional mean
may fail to appropriately recognize potential heterogeneity, i.e., the predictive value of a particular variable in the center
of the conditional distribution of a response can be very different from that, say in the
tails. By estimating quantiles at different parts of the conditional distribution, a more complete picture of the relationship between variables and the response can be obtained.

To identify informative variables for the purpose of estimating conditional quantiles, 
many penalized quantile regression approaches have been proposed. For example, 
\cite{belloni2011} considered quantile regression in high dimensional setting using the LASSO penalty~\citep{tibshirani1996regression}. 
\cite{wangwuli2012} and \cite{sherwood2016} investigated the penalized quantile regression with ultra-high dimensionality using a
non-convex penalty function, including the SCAD~\citep{fan2001variable} and the MCP~\citep{zhang2010nearly}. 
These penalized approaches can select important variables and estimate parameters simultaneously, and also enjoy nice theoretical properties.  However, in the ultra-high dimensional setting where the dimension grows exponentially with the sample size, 
penalized approaches can suffer from three practical issues, including computational cost, statistical accuracy, and model
interpretability~\citep{fan2009ultrahigh, fan2010selective}. 

To address these issues, a natural idea is to reduce the dimensionality. Motivated by the sure independence screening method for linear regression in~\cite{fanlv2008}, various 
marginal screening methods have been proposed under the quantile regression framework; see, for example, \cite{he2013} and \cite{wuyin2015}. Although these marginal screening methods are
computationally efficient, they may miss some relevant variables that truly contribute to the quantiles conditionally but not marginally.
In addition, marginal screening methods tend to select some irrelevant covariates when the variables are highly correlated.
To handle these problems, \cite{maetal2017}  
used forward regression methods for variable screening in the linear quantile regression setting. 
The main idea in \cite{maetal2017} is to select relevant covariates in a forward regression manner. 
Such an idea was first introduced by~\cite{wang2009} for linear regression and has been extended to other models in the mean regression framework; see, for example, \cite{cheng2016forward},  \cite{li2017profile}, \cite{cheng2018greedy},  
\cite{zhong2020forward}, and references therein. 

However, in practice, there is often little prior information suggesting that the effects of the covariates on the conditional quantile take a linear form. There is a clear need to relax the parametric assumption posed  
in \cite{maetal2017}. 
In this paper, we consider
nonparametric additive quantile regression models and develop a new 
variable screening
procedure in the ultra-high dimensional setting.  
We first propose an additive quantile forward screening (AQFS) algorithm for variable screening, which will be introduced in Section~\ref{sec: screening}. 
The proposed AQFS algorithm selects relevant covariates sequentially in a forward regression manner and adds one new covariate into the selected set in each step. We show that our algorithm can
identify all relevant covariates in finite steps. To further remove noise covariates after the screening stage, we  apply a modified quantile Bayesian information criterion (QBIC) for best subset selection. We prove that the resulting best subset still contains all the relevant covariates with overwhelming probability.

The main contributions of this paper are as follows. First, we propose a new variable screening  
procedure for ultra-high dimensional additive quantile regression models, which has been proved to enjoy the sure screening property in the sense that all relevant covariates can be selected with probability approaching one. To the best of our knowledge, this is the first one to consider forward regression for variable screening in the ultra-high dimensional additive quantile regression setting.  
Second, the proposed AQFS algorithm extends the screening method for linear regression in \cite{wang2009} and that for linear quantile models in \cite{maetal2017} to nonparametric additive quantile regression. Compared to the existing method in
\cite{maetal2017}, our procedure  
is more flexible and more robust to model misspecification. 
This makes our procedure more applicable in real applications.
Third, by using the QBIC criterion,
we provides a non-penalized variable selection method to further remove noise covariates in the screened set. Under some mild conditions, we show that the resulting best subset also enjoys the sure screening property. Fourth, our procedure is computationally fast to be implemented. This is very appealing for practitioners in the ultra-high dimensional setting.

The rest of the paper is organized as follows.  Section~\ref{sec:method} 
provides a detailed account of the proposed additive quantile forward regression for variable screening and selection when faced with ultra-high dimensional data. Section~\ref{sec: theory} establishes the theoretical properties of our proposed screening and selection methods. 
Sections~\ref{sec:simulation} and~\ref{sec:application} demonstrate the advantages and numerical performances of our procedure 
through simulation studies and a real data example, respectively.  All the proofs, technical details, and additional simulation results are relegated to the Appendix.

\section{The Proposed Methodology}\label{sec:method}

We consider the nonparametric additive quantile model, where the $\tau$th ($0 < \tau < 1$) conditional quantile of the outcome variable $Y$ given a $p$-dimensional vector of covariates $\bx=(X_1, \dots, X_p)^\top$ is
\begin{eqnarray}\label{eq: AQR-model}
Q_{\tau}(Y|\bx) = \mu_\tau+ \sum^p_{j=1} g_{j,\, \tau}(X_{j}),
\end{eqnarray}
in which $\mu_\tau$ is a constant and $g_{j,\, \tau}(X_j), j=1, \dots, p, $ are smooth nonparametric functions. For identification, it will be assumed that ${\rm E}(g_{j,\, \tau}(X_j))=0$. 
Such an additive quantile regression model was also considered by~\cite{de2003additive}, \cite{Horowitz2005nonparametric}, and~\cite{cheng2011efficient} with the focus on the nonparametric estimation of each unknown function $g_{j,\, \tau}(\cdot)$ for a fixed dimension $p$. 
In this paper, we consider feature screening for model \eqref{eq: AQR-model} in ultra-high dimensional setting where the dimensionality $p$ can grow exponentially with the number of observations $n$. 
We denote the set of relevant  
(or important) covariates by
\begin{eqnarray}\label{eq:true}
\mathcal{S}_{\tau}^*=  \Bigl \{1\leq j \leq p: \exists\, x_j \in \mathcal{X}_j,~ g_{j,\, \tau}(x_{j}) \neq 0 \Bigr  \},
\end{eqnarray}
where $\mathcal{X}_j$ is the support of $X_j$. Let $p^*_\tau$ be the number of important covariates, i.e. 
$|\mathcal{S}_{\tau}^*|=p^*_\tau$ where $|A|$ denotes the number of elements in a set $A$. We assume  quantile effect sparsity in the sense that at each $\tau$ only a small number of covariates ($p^*_\tau \ll p$) are actively associated with the $\tau$th conditional quantile of $Y$. By considering
several $\tau$'s , one may be able to detect a large number of important covariates that truly influence the various parts of the conditional distribution of $Y$. To simplify the presentation, we drop the subscript $\tau$ in $\mu_\tau$,  $g_{j, \tau}$, $\mathcal{S}_{\tau}^*$ and $p^*_\tau$ for the remainder of the paper  when there is no confusion.

\subsection{Covariate Importance Ranking \label{sec: Variable Importance} }

Let $ \mathcal{F}=\{1, \ldots, p\}$.
We use a generic notation $\mathcal{S} = \{j_1, \cdots, j_{s} \} \subseteq \mathcal{F}$ to denote an arbitrary additive conditional quantile model with size (or cardinality) $|\mathcal{S}|$. Although model $\mathcal{S}$ may vary with $\tau$, we drop $\tau$ for notational convenience. Mimicking (\ref{eq: AQR-model}), the $\tau$th conditional quantile of $Y$ given $\bx_{\mathcal{S}}=\{X_j: j\in\mathcal{S}\}$  
is
\begin{eqnarray}\label{eq:AddS}
Q_{\tau}(Y|\bx_{\mathcal{S}}) = \mu_{\mathcal{S}} + \sum_{j \in \mathcal{S}} g_j^{\mathcal{S}}(X_j),
\end{eqnarray}
where $\mu_{\mathcal{S}}$ is a constant and  $g_j^{\mathcal{S}}(\cdot)$, $j\in S$, are smooth nonparametric functions 
with ${\rm E}(g_j^{\mathcal{S}}(X_j))=0$.
Note that $\mu_{\mathcal{S}}$ may vary with $\mathcal{S}$. For example, 
when $\mathcal{S}=\emptyset$, 
$Q_{\tau}(Y|\emptyset)=Q_{\tau}(Y)$, which is the $\tau$-th unconditional quantile of $Y$.

The B-spline approximation technique has been widely used to approximate unknown functions in nonparametric regression; see, for example, \cite{sherwood2016}, \cite{fan2011}, \cite{he2013},
\cite{cheng2016forward},
\cite{jiang2020functional}, 
and references therein.
Following these existing works, we also use B-spline basis functions to estimate the unknown  $g_j^{\mathcal{S}}(\cdot)$ in \eqref{eq:AddS} in our paper.
Without loss of generality, we assume that each $X_j$ takes values on the interval $[0, 1]$. Let $\bB_j(t)= (B_{j,\,1}(t), \dots, B_{j,\,q_n}(t))^\top$ denote a vector of normalized B-spline basis functions of order $\ell + 1$ with $k_n$ knots, where the number of basis functions $q_n=k_n+\ell$, $\|B_{j, \,k}\|_{\infty}\leq 1$ for each $1\leq k\leq q_n$, and $\|\cdot\|_{\infty}$ denotes the sup norm. 
Throughout this paper, we choose $q_n=O\left(n^{1/(2d+1)}\right)$, 
which is the optimal convergence rate for B-spline approximation in nonparametric regression~\citep{stone1985additive},
where 
$d>1$ is specified later in Section~\ref{sec: theory}.
Under some smoothness conditions, the nonparametric functions  $g_j^{\mathcal{S}}(X_j)$ can be well approximated by a linear
combination of the basis functions, $\bB_j^\top(X_j) \boldsymbol{\theta}^{\mathcal{S}}_j$, for some $\boldsymbol{\theta}^{\mathcal{S}}_j \in \mathbb{R}^{q_n}$~\citep{stone1985additive}, that is, 
\begin{align}\label{eq:bspline}
g_j^{\mathcal{S}}(X_j)\approx \bB_j^\top(X_j) \boldsymbol{\theta}^{\mathcal{S}}_j.
\end{align}

Write  $\boldsymbol{\theta}_{\mathcal{S}}=(\mu_{\mathcal{S}}, \boldsymbol{\theta}^{\mathcal{S}}_1, \dots, \boldsymbol{\theta}^{\mathcal{S}}_{|\mathcal{S}|})$ and 
$\bz_{\mathcal{S}}=(1, \bPi^{\top}_{\mathcal{S}})^{\top}$, where 
$\bPi_{\mathcal{S}}$ is the column vector formed by stacking vectors $\bB_j(X_j)$, $j\in\mathcal{S}$.
Then, using (\ref{eq:bspline}), 
the $\tau$th conditional quantile of $Y$ given $\bx_{\mathcal{S}}$ in (\ref{eq:AddS}) can be approximated by
\begin{eqnarray}\label{eq: Q_approx}
Q_{\tau}(Y|\bx_{\mathcal{S}}) 
\approx \mu_{\mathcal{S}} + \sum_{j \in \mathcal{S}} \bB_j^\top (X_j)\boldsymbol{\theta}^S_j
=\bz_{\mathcal{S}}^T\boldsymbol{\theta}_{\mathcal{S}},
\end{eqnarray}
where the only unknown is the $N_{\mathcal{S}}$-dimensional parameter vector  $\boldsymbol{\theta}_{\mathcal{S}}$ 
with
$N_\mathcal{S} =1 + q_n|\mathcal{S}|$. 
In practice, given the observed data $\{(\bx_i^{\top}, Y_i), i=1,\cdots, n\}$, which are $n$ independent and identically distributed copies of  $(\bx^{\top}, Y)$, we can estimate $\boldsymbol{\theta}_{\mathcal{S}}$ 
through
\begin{eqnarray*}
	\boldsymbol{\hat{\theta}}_{\mathcal{S}} = \arg\min_{\boldsymbol{\theta}_{\mathcal{S}} } \sum^n_{i=1} \rho_\tau \left (Y_i - \bz_{i,\,\mathcal{S}}^T\boldsymbol{\theta}_{\mathcal{S}}\right) 
\end{eqnarray*}
where $\rho_{\tau}(u)=u[\tau-I(u<0)]$ is the quantile check function and $\bz_{i, \,\mathcal{S}}=(1, \bPi^\top_{i,\,\mathcal{S}})$ for each $i=1, \dots, n$ with $\bPi_{i,\,\mathcal{S}}$ being the column vector formed by stacking vectors $\bB_j(X_{ij})$, $j\in\mathcal{S}$. Then, an estimate of (\ref{eq: Q_approx}) 
is 
\begin{eqnarray}\label{eq: Q_approxE}
\widehat{Q}_{\tau}(Y|\bx_{i,\,\mathcal{S}})=\bz_{i,\,\mathcal{S}}^T \boldsymbol{\hat{\theta}_{\mathcal{S}}}.
\end{eqnarray}

Next, we introduce our conditional importance score, which can be used to rank the importance of covariates. For any $k \in \mathcal{F} \backslash \mathcal{S}$, 
consider a candidate covariate $X_k$. 
If the $\tau$th conditional quantile of $Y$ given $\bx_{\mathcal{S}\cup \{k\}}$ is not associated with $X_k$, it will follow that  $Q_{\tau}(Y|\bx_{\mathcal{S}\cup \{k\}})=Q_{\tau}(Y|\bx_{\mathcal{S}})$. Using this, it can be seen that    
\begin{align}\label{eq: Cond-xM-Xk}
&{\rm E}\left[\tau - I\{Y - Q_{\tau}(Y | \bx_{\mathcal{S}}) < 0 \} | X_k\right] 
=  {\rm E}\left[\tau - I\{Y - Q_{\tau}(Y | \bx_{\mathcal{S}\cup \{k\}}) < 0 \} | X_k\right] \nonumber\\
=& {\rm E}\left\{{\rm E}[\tau - I\{Y - Q_{\tau}(Y|\bx_{\mathcal{S}\cup \{k\}}) < 0 \} |\bx_{\mathcal{S}\cup \{k\}}]\big| X_k \right\}
={\rm E}(0|X_k)=0,
\end{align}
where $I(\cdot)$ is the indicator function, the second equality comes from the property of conditional expectation, and the third equality follows from the definition of conditional quantile.
Motivated by this observation, 
we introduce  
the conditional  importance score 
for the $k$th covariate
\begin{align}\label{eq: dk-def}
d_k(t|\mathcal{S})={\rm E}\Big(\left[\tau - I\{Y < Q_{\tau}(Y | \bx_{\mathcal{S}})  \}\right]I(X_k<t)\Big).
\end{align}
If $X_k$ is unimportant given model $\mathcal{S}$, 
it can be seen from $\eqref{eq: Cond-xM-Xk}$ that 
$d_k(t|\mathcal{S})=0$ 
for any $t$ in the support of $X_k$. 
More generally, one can replace $I(X_k<t)$ in \eqref{eq: dk-def} by $f(X_k, t)$, some function of $X_k$ and 
$t$ to construct $d_k(t|\mathcal{S})$.
In this paper we use $I(X_k<t)$ because the approach is based on the indicator function and does not require a finite-moment assumption for each $X_k$. Thus, it offers more robustness with respect to heavy-tailed distributions.

Subsequently, we define the sample version of the conditional covariate importance score in (\ref{eq: dk-def}) as
\begin{align}\label{eq: dkt-est}
\widehat{d}_k(t|\mathcal{S})
=n^{-1}\sum_{i=1}^n
\left[\tau - I\{Y_i < \widehat{Q}_{\tau}(Y|\bx_{i,\,\mathcal{S}})\}\right]I(X_{ik}<t),
\end{align}
where $\widehat{Q}_{\tau}(Y|\bx_{i,\,\mathcal{S}})$ is given in~\eqref{eq: Q_approxE}. 
If the $\tau$th conditional quantile of $Y$ given $\bx_{\mathcal{S}\cup \{k\}}$ is not associated with $X_k$ (that is, 
candidate covariate $X_k$ is conditionally unimportant given baseline 
model $\mathcal{S}$), we expect 
$\widehat{d}_k(t|\mathcal{S}) \approx 0$ for any $t$ in the support of $X_k$.
We further define
\begin{align}\label{eq: dk}
\|\widehat{d}_k(\mathcal{S})\|
=n^{-1}\sum_{i=1}^n\hat{d}_k^2(X_{ik}|\mathcal{S}).
\end{align}
Following the rationale for (\ref{eq: dkt-est}), we can use  $\|\widehat{d}_k(\mathcal{S})\|$ to rank candidate covariates by importance given a baseline model $\mathcal{S}$. In particular, covariates with a large value of $\|\widehat{d}_k(\mathcal{S})\|$ are considered relatively more important.

\subsection{The AQFS Screening Algorithm \label{sec: screening} }
 
We now introduce a new variable screening algorithm aimed at identifying the true model $\mathcal{S}^*_{\tau}$ in (\ref{eq:true}). The main idea of our screening algorithm is to select variables sequentially by finding the variable with the maximal sample 
conditional covariate importance score
on the current selected active set
and then
adding it to the selected active set in each step. 
More specifically, denote by 
$\mathcal{S}^{(\ell)}$ the set of retained variables in the $\ell$ forward screening step. 
We start with $\mathcal{S}^{(0)}=\emptyset$,  compute 
$\|\widehat{d}_k(\mathcal{S}^{(0)})\|$ using (10) with $\mathcal{S}=\mathcal{S}^{(0)}$ for each $k\in\{1, \cdots, p\}\backslash \mathcal{S}^{(0)}$, and find the variable index $a_{1}$ that has the maximal sample conditional importance score, i.e., $a_{1}=\argmax_{k\in\{1, \cdots, p\}}\,\|\widehat{d}_k(\mathcal{S}^{(0)}))\|$, and obtain 
$\mathcal{S}^{(1)}=\mathcal{S}^{(0)}\cup\{a_{1}\}=\{a_{1}\}$. Then, we move to the next step and compute 
$\|\widehat{d}_k(\mathcal{S}^{(1)})\|$ using (10) with $\mathcal{S}=\mathcal{S}^{(1)}$ for each $k\in\{1, \cdots, p\}\backslash \mathcal{S}^{(1)}=\{1, \cdots, p\}\backslash \{a_1\}$, find the variable index $a_{2}$ that has the maximal sample conditional importance score, i.e., $a_{2}=\argmax_{k\in\{1, \cdots, p\}\backslash \mathcal{S}^{(1)}}\,\|\widehat{d}_k(\mathcal{S}^{(1)}))\|$, and obtain 
$\mathcal{S}^{(2)}=\mathcal{S}^{(1)}\cup\{a_{2}\}=\{a_{1}, a_{2}\}$. We repeat the above step $K_n$ times, resulting in a total of $K_n$ nested models $\mathcal{S}^{(1)}\subset \mathcal{S}^{(2)}\subset \cdots \subset \mathcal{S}^{(K_n)}$, with $\mathcal{S}^{(\ell)}=\{a_1, \cdots, a_{\ell}\}$. Our screening procedure, named AQFS, 
is summarized below in Algorithm~\ref{algorithm:AQFR}.

\begin{algorithm}[H]
	\caption{AQFS}
	\label{algorithm:AQFR}
	\begin{algorithmic}[0]
		\State \textbf{Step 1} (Initialization). Set $\mathcal{S}^{(0)}=\emptyset$ and $\mathcal{F} = \{1, \dots, p \}$.
		\State \textbf{Step 2} (Forward Addition).
		\begin{itemize}
			\setlength{\itemsep}{0pt}	
			\item[(a)] {\it Evaluation}. In the $\ell$th step ($\ell\geq 1$), we are given $\mathcal{S}^{(\ell-1)}$.  
			Then, for each $k \in \mathcal{F} \backslash \mathcal{S}^{(\ell-1)}$, we compute $\|\widehat{d}_k(\mathcal{S}^{(\ell-1)}))\|$ using \eqref{eq: dk} with $\mathcal{S}=\mathcal{S}^{(\ell-1)}$.
			\item[(b)] {\it Screening}. Screen the sample conditional importance score for all $k \in \mathcal{F} \backslash \mathcal{S}^{(\ell-1)}$ to find the variable index $a_{\ell}$ that has the maximal sample conditional importance score, i.e., $a_{\ell}=\argmax_{k \in \mathcal{F} \backslash \mathcal{S}^{(\ell-1)}}\,\|\widehat{d}_k(\mathcal{S}^{(\ell-1)}))\|$.
			\item[(c)] {\it Updating}. Let $\mathcal{S}^{(\ell)}=\mathcal{S}^{(\ell-1)}\cup\{a_{\ell}\}$.		  
		\end{itemize}
		\State \textbf{Step 3}  (Solution Path). Repeat Step 2 $K_n$ times, resulting in a total of $K_n$ nested models $\mathcal{S}^{(1)}\subset \mathcal{S}^{(2)}\subset \cdots \subset \mathcal{S}^{(K_n)}$.      
	\end{algorithmic}
\end{algorithm}

In practice, we can take $K_n=[n/\log(n)]$ when implementing our screening algorithm since this order can guarantee the sure screening property meaning that all of the relevant covariates can be retained with probability tending to one; see Remark 3 in Section~\ref{sec: theory}.  Throughout this paper, we use $[b]$ to denote the integer part of a number $b$. 
Such a choice of $K_n$ results in the estimated active set containing $K_n=[ n/\log (n)]$ covariates after the screening stage. 
The same choice for the number of covariates kept in the screening stage is 
also widely used in the variable screening literature; see, for example, \cite{fanlv2008}, \cite{fan2011},  \cite{li2012feature}, \cite{he2013},  \cite{maetal2017}, 
and~\cite{zhong2023feature}.
 
When $\mathcal{S}=\emptyset$, 
our conditional covariates importance score $d_{k}(t|\mathcal{S})$ given in \eqref{eq: dk-def} reduces to 
the unconditional importance score $d_k(t)={\rm E}\Big(\left[\tau - I\{Y < Q_{\tau}(Y)\}\right]I(X_k<t)\Big)$ in 
~\cite{wuyin2015}. In this sense, our conditional covariate importance score $d_{k}(t|\mathcal{S})$  can be seen as an extension of the unconditional importance score of~\cite{wuyin2015}.  However, unlike the screening approach of ~\cite{wuyin2015} to select all important variables via a one-time hard-thresholding through $d_k(t)$, we adopt the conditional covariate importance score $d_{k}(t|\mathcal{S})$ to screen potentially important variables in a forward regression manner by iteratively updating  the baseline model $\mathcal{S}$.
In addition, \cite{maetal2017} proposed subsequently selecting relevant covariates for linear quantile regression models via quantile partial correlation.   
Our conditional covariate importance score $d_{k}(t|\mathcal{S})$ is different from the quantile partial correlation in \cite{maetal2017}. On the other hand, there is often little prior information (especially in high dimensions) in practice indicating whether the influence of  covariates is linear or not. 
Therefore, our conditional covariate importance score $d_{k}(t|\mathcal{S})$ offers more flexibility over the covariate importance scores in \cite{wuyin2015} and \cite{maetal2017}.

\subsection{ Selection by QBIC \label{sec: selection} }

Applying the AQFS algorithm, we can reduce the ultra-high dimension $p$ to a more manageable size $K_n$ via variable screening. Note that the AQFS algorithm has generated $K_n$ nested models $\mathcal{S}^{(1)}\subset \mathcal{S}^{(2)}\subset \cdots \subset \mathcal{S}^{(K_n)}$. To find the best model among these $K_n$ models, by following~\cite{LeeNohPark2014} and using (\ref{eq: Q_approxE}), we consider the following modified Bayesian information criterion for our additive quantile regression (QBIC)
\begin{eqnarray}\label{eq: QBIC}
\mbox{QBIC}(\mathcal{S})
=\log \Biggl [\sum_{i=1}^n 
\rho_{\tau} \Bigl (Y_i- \widehat{Q}_{\tau}(Y|\bx_{i,\,\mathcal{S}}) \Bigr ) \Biggr]
+ N_{\mathcal{S}}  \frac{\log n}{2n} C_n,
\end{eqnarray}
where $C_n$ can depend on $n$ and $p$ only and is allowed to diverge with the sample size $n$. 
Throughout this paper, we use the notation $C_n$ instead of $C_{n, \, p}$ for brevity since we consider model~\eqref{eq: AQR-model} in the ultra-high dimensional setting where the number of covariates $p$ can grow exponentially with the number of observations $n$. The diverging order of $C_n$ will be specified in 
Theorem~\ref{Th4} of Section \ref{sec: theory}.
This QBIC criterion extends ordinary BIC and is suitable for model selection in large model spaces. Let  
\begin{align}\label{eq: ell_hat_def}
\hat{\ell}=\argmin_{1\leq \ell\leq K_n}\mbox{QBIC}(\mathcal{S}^{(\ell)})
\end{align}
and denote the resulting best selected model as $\widehat{\mathcal{S}}_{\mathrm{QBIC}}=\mathcal{S}^{(\hat{\ell})}$.
It will be shown in Section~\ref{sec: theory} that 
the reduced model $\widehat{\mathcal{S}}_{\mathrm{QBIC}}$ also enjoys the sure screening property, i.e.,
$P(\mathcal{S}^{*} \subset  \widehat{\mathcal{S}}_{\mathrm{QBIC}})\rightarrow 1$ as $n\to\infty$.

\subsection{Illustration of Our Method}
Our method is a two-stage procedure: 1) screening using AQFS and 2) selection via QBIC. To be more specific, we first employ our variable screening algorithm AQFS to select variables sequentially and obtain a total of $K_n$ nested models $\mathcal{S}^{(1)}\subset \mathcal{S}^{(2)}\subset \cdots \subset \mathcal{S}^{(K_n)}$.  Then we apply the QBIC criterion in~\eqref{eq: QBIC} for best subset selection (i.e., use the QBIC to choose the best model among these $K_n$ nested models).
Of course, if one is only interested in variable screening, the outcome of the proposed method is the set $\mathcal{S}^{(K_n)}$ consisting of $K_n$ variables.

\begin{figure}[H]
	\centering
	\includegraphics[width=0.9\textwidth]{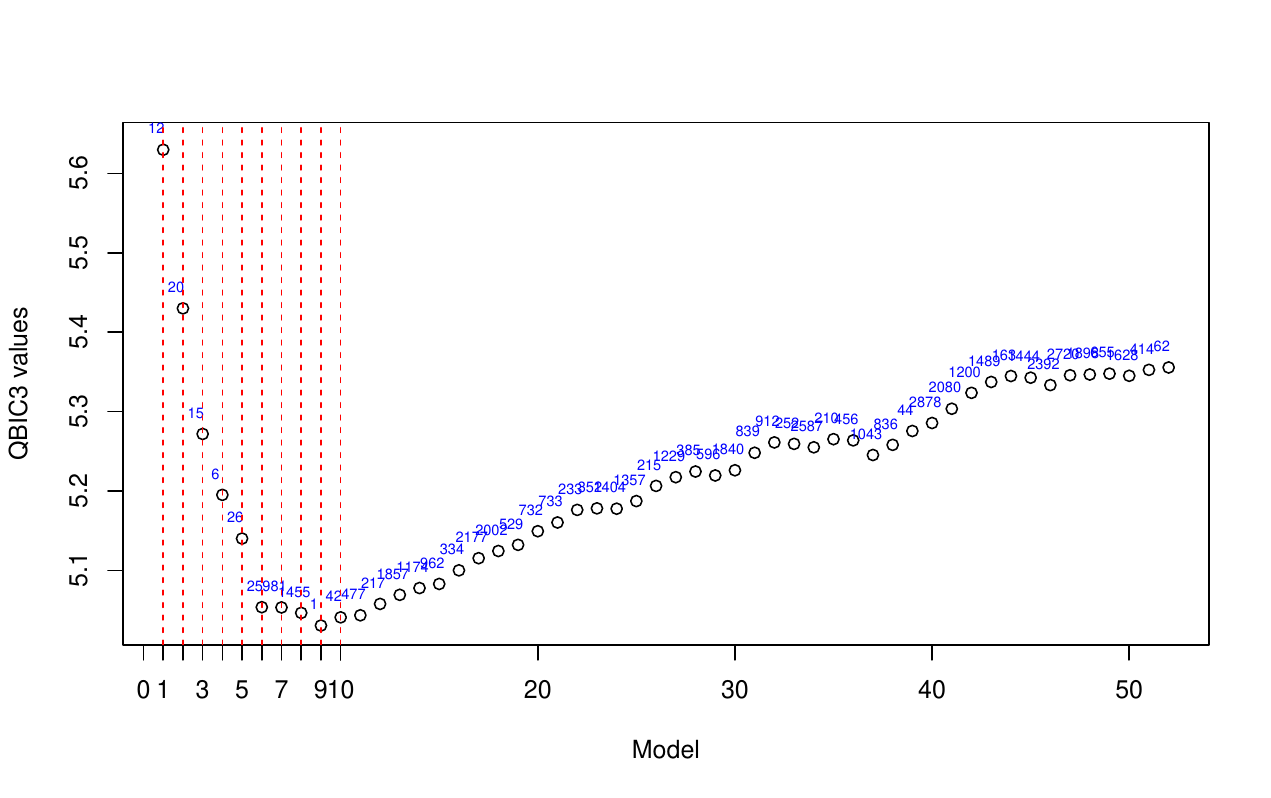}
	\vspace*{-8mm}
	\caption{Illustration of our method with $(n, p)=(300, 3000)$.}
	\label{fig:rep1-qbic3-new}
\end{figure}

To demonstrate our method, consider the model 
\begin{align*}
Y = \beta_1 X_6 +  \beta_2 X_{12} + \beta_3 X_{15} + \beta_4 X_{20} + g_{25}(X_{25}) + g_{26}(X_{26}) + 0.7 X_1 \varepsilon
\end{align*}
where truly relevant covariates at $\tau=0.7$ are $X_1, X_6, X_{12}, X_{15}, X_{20}, X_{25}, X_{26}$ (i.e., $\mathcal{S}^*_\tau=\{1, 6, 12, 15, 20, 25, 26\}$ at $\tau=0.7$) and $(n, p)=(300, 3000)$.  We simulated data in the
same way as that for Example 2 in Section~\ref{sec:simulation}.
Figure~\ref{fig:rep1-qbic3-new} shows that the screening and selection of our method with QBIC3 (i.e., $C_n=\log(\log p^{0.5})$ in~\eqref{eq: QBIC}; see Section~\ref{sec:simulation} for details) for one replication.
The newly identified variable in each step of the screening stage is marked in blue for clarity, and red dashed vertical lines are used to more clearly indicate the step at which each variable is identified.
As shown in Figure~\ref{fig:rep1-qbic3-new}, we started with $\mathcal{S}^{(0)}=\emptyset$ and 
found $a_{1}=\argmax_{k\in\{1, \cdots, p\}}\,\|\widehat{d}_k(\mathcal{S}^{(0)}))\|=12$, meaning that the variable $X_{12}$ was firstly identified and $\mathcal{S}^{(1)}=\{12\}$. 
Then, we moved to the next step, computed 
$\|\widehat{d}_k(\mathcal{S}^{(1)})\|$ using (10) with $\mathcal{S}=\mathcal{S}^{(1)}$ in our paper for each $k\in\{1, \cdots, p\}\backslash \mathcal{S}^{(1)}=\{1, \cdots, p\}\backslash \{12\}$, and found $a_{2}=\argmax_{k\in\{1, \cdots, p\}\backslash \mathcal{S}^{(1)}}\,\|\widehat{d}_k(\mathcal{S}^{(1)}))\|=20$. So the variable $X_{20}$ was identified in the second step of the screening stage and 
$\mathcal{S}^{(2)}=\mathcal{S}^{(1)}\cup\{a_{2}\}=\{12, 20\}$.  
Similarly, variables $X_{15}$ and $X_{6}$ were identified, respectively, during the third and fourth steps of the screening stage. In other words, $a_3={15}$, $a_4=6$, 
$\mathcal{S}^{(3)}=\{12, 20, 15\}$, and $\mathcal{S}^{(4)}=\{12, 20, 15, 6\}$.
After $K_n=[n/\log(n)]=52$ steps in the screening stage, we obtained 52 nested models $\mathcal{S}^{(1)}\subset \mathcal{S}^{(2)}\subset \cdots \subset \mathcal{S}^{(52)}$ and the set $\mathcal{S}^{(52)}$ includes the variable index for all the truly relevant covariates.
Then we used QBIC to choose the best model among these nested models.  It can be seen that the best model chosen by QBIC3 is
$\mathcal{S}^{(9)}=\{12, 20, 15, 6, 26, 25, 981, 1455, 1\}$.

\vspace{-10mm}

\section{Theoretical Properties }\label{sec: theory}

In this section, we establish the sure independent screening properties of the AQFS screened set $\mathcal{S}^{(K_n)}$ and the subsequent QBIC-selected set    
$\widehat{\mathcal{S}}_{\mathrm{QBIC}}$.  Let $r_n$ be an upper bound on the cardinality $|\mathcal{S}|$ 
and $r_n$ is allowed to increase with the sample size $n$. The key step in deriving the sure screening properties is to establish the uniform convergence of the empirical covariate importance ranking score  $\|\widehat{d}_{k}(\mathcal{S})\|$ 
(\ref{eq: dk})
to its population counterpart, $\|d_{k}(\mathcal{S})\|={\rm E}[d^2_k(t|\mathcal{S})]$, where $d_k(t|\mathcal{S})$ is as defined in \eqref{eq: dk-def}. The uniform convergence result is formally stated in Theorem~\ref{Th1}.
We impose the following regularity conditions to facilitate our technical analysis.

\begin{condition}\label{con: Lipschitz}
	The nonparametric components $\left\{g_j^{\mathcal{S}}(\cdot), j\in\mathcal{S}\subseteq\{1, \cdots, p\}\right\}$ belong to $\mathbb{F}$, where $\mathbb{F}$ is the class of functions whose $m$th derivative satisfies a Lipschitz condition of order $c_1$. That is,
   $\mathbb{F}=\left\{f(\cdot): |f^{(m)}(t_1)-f^{(m)}(t_2)|\leq L|t_1-t_2|^{c_1}\,\,\mbox{for all}\,\, t_1, t_2\in[0, 1]\right\}$
	for some positive constant $L$, where $m$ is a nonnegative integer and $0<c_1\leq 1$ such that $d=m+c_1>1$.
\end{condition}

\begin{condition}\label{con: fy}	
	Let $F_{Y|\bx}(y)$ and $f_{Y|\bx}(y)$ are the conditional distribution and density functions of $Y$ given $\bx$, respectively. In a neighborhood of $Q_{\tau}(Y|\bx_{S})$, $F_{Y|\bx}(y)$ is twice differentiable, $f_{Y|\bx}(y)$ is uniformly bounded away from $0$ and $\infty$, and $f'_{Y|\bx}(y)$, the derivative of $f_{Y|\bx}(y)$, is uniformly bounded.  	
\end{condition}

Condition~\ref{con: Lipschitz} is the
Lipschitz assumption requiring each nonparametric function $g_j^{\mathcal{S}}$ to belong to a class of smooth functions. 
This condition is often used in the literature; See, for example,  \cite{huang2010variable}, \cite{fan2011}, \cite{he2013}, \cite{zhong2020forward}, and references therein.  
\cite{stone1985additive}
showed that functions
satisfying the Lipschitz condition can be effectively approximated by B-splines basis functions. Condition~\ref{con: fy}	is standard in the literature on quantile regression; see, for example, \cite{wuyin2015}.

\begin{theorem}\label{Th1} 
	Under Conditions \ref{con: Lipschitz}-\ref{con: fy}
	and for any constant $C >0 $, there exist some positive constants $c_2$ and $c_3$ such that, for $0<\alpha<d/(2d+1)$,  
	$r_n = O(n^{\omega})$ with $0\leq \omega< \min\{2\alpha,\, 2d/(2d+1)-2\alpha\}$ and $q_n=O\left(n^{1/(2d+1)}\right)$, we have
	\begin{align*}
	P\left(\max_{1\leq k\leq p}\big|\|\widehat{d}_{k}(\mathcal{S})\|-\|d_{k}(\mathcal{S})\|\big|\geq 5Cr_n^{1/2}n^{-\alpha}\right)
	\leq  2p\exp\left(-c_2r_n^2n^{1-4\alpha}\right)+4p\exp\left(-c_3r_nn^{1-2\alpha}\right)
	\end{align*}
	for all $n$ sufficiently large.
\end{theorem}

\noindent{\bf Remark 1}. Recall that we choose $q_n=O\left(n^{1/(2d+1)}\right)$ throughout our paper, 
which is the optimal convergence rate for B-spline approximation in nonparametric regression~\citep{stone1985additive}. 
When we take $r_n=O(n^{\omega})$, the requirement $\omega<2d/(2d+1)-2\alpha$ for  $0<\alpha<d/(2d+1)$ is equivalent to the assumption $r_n^{1/2}q_n^{-d}n^{\alpha}=o(1)$. In this sense, the requirement $\omega<2d/(2d+1)-2\alpha$ for  $0<\alpha<d/(2d+1)$ can be understood as a constraint on how fast the number of basis function $q_n$ is allowed to increase with the sample size $n$. Similar condition on $q_n$ is also used in~\cite{fan2011} and~\cite{he2013}.

\medskip
\noindent{\bf Remark 2}.  Theorem \ref{Th1} demonstrates the uniform convergence of $\|\widehat{d}_{k}(\mathcal{S})\|$   to its population counterpart. This result   
indicates that we can handle 
ultra-high dimensional data with  
$\log p = o\left(r_n^2n^{1-4\alpha} + r_nn^{1-2\alpha}\right)$. Hence, 
$p$ can grow with the sample size $n$ at an exponential rate.

\medskip

To study the sure screening property of our approach, we assume the following condition on signal strength for all relevant covariates in $\mathcal{S}^*$.

\begin{condition}\label{con: min-signal}
	$\min_{k\in \mathcal{S}*} \|d_{k}(\mathcal{S})\|\geq 2C_0r_n^{1/2}n^{-\alpha}$
	for some positive constant $C_0$.
\end{condition}

Condition~\ref{con: min-signal} is an assumption on the minimum signal strength for all relevant covariates. A smaller constant $\alpha$ corresponds to a stronger signal. This condition is important as it guarantees that the proposed covariate importance ranking score carries information about the relevant variables. Similar conditions are used in the variable screening literature; see, for example, 
\cite{fanlv2008}, \cite{li2012feature}, \cite{he2013}, \cite{song2014censored}, \cite{maetal2017},  \cite{kong2017},  
\cite{Li2022high}, and references therein.

Recall that our proposed variable screening approach, AQFS, selects covariates sequentially by finding the covariate with the maximal sample  
covariate importance ranking score
and then adding it to the selected relevant set in each step.  To facilitate the theoretical analysis, we use 
$\widehat{\mathcal{S}}_{v_n}$ to denote the resulting relevant set via our AQFS algorithm such that 
the sample   
covariate importance ranking scores of the selected covariates in $\widehat{\mathcal{S}}_{v_n}$ are greater than or equal to a threshold $v_n$, that is, 
\begin{eqnarray}\label{eq: S-vn}
\widehat{\mathcal{S}}_{v_n} = \left\{k: \|\widehat{d}_{k}(\mathcal{S}^{(\ell -1)})\| \geq v_n\,\,\mbox{for}\,\,1\leq \ell\leq K_n \right\}.
\end{eqnarray} 
The following theorem shows the sure screening property that all relevant covariates can be retained in  $\widehat{\mathcal{S}}_{v_n}$ with probability tending to one.

\begin{theorem}\label{Th2}
	Take $v_n=C_0r^{1/2}_nn^{-\alpha}$. For $0<\alpha<d/(2d+1)$,  
	$r_n = O(n^{\omega})$ with $0\leq \omega< \min\{2\alpha,\, 2d/(2d+1)-2\alpha\}$ and $q_n=O\left(n^{1/(2d+1)}\right)$, under the conditions in Theorem~\ref{Th1} and Condition~\ref{con: min-signal}, we have 
	\begin{align*}
	P(\mathcal{S}^{*} \subset  \widehat{\mathcal{S}}_{v_n})\geq 1-|\mathcal{S}^{*}|\left[2\exp\left(-c_2 r^2_n n^{1-4\alpha}\right)+2\exp\left(-c_3 r_n n^{1-2\alpha}\right)\right]
	\end{align*}
	for all $n$ sufficiently large.
\end{theorem}

Theorem~\ref{Th2} reveals that the probability bound for the sure screening property depends on the number of relevant covariates (i.e., $|\mathcal{S}^{*}|$), but not on the number of covariates $p$.  
It can be seen that when the signal strength for relevant covariates becomes stronger, the sure screening property holds for higher dimensionality $p$.
Another interesting question is how many covariates are retained after the screening. To control the size of the selected model, we assume 
$\sum\limits_{k=1}^p\|d_{k}(\mathcal{S})\|=O(n^{\eta})$ for some constant $\eta>0$.
The following theorem provides an upper bound on the size of selected covariates after screening.

\begin{theorem}\label{Th: size}
	Take $v_n=C_0r^{1/2}_nn^{-\alpha}$ and assume $\sum\limits_{k=1}^p\|d_{k}(\mathcal{S})\|=O(n^{\eta})$ for some $\eta>0$. Then for $0<\alpha<d/(2d+1)$,  
	$r_n = O(n^{\omega})$ with $0\leq \omega< \min\{2\alpha,\, 2d/(2d+1)-2\alpha\}$ and $q_n=O\left(n^{1/(2d+1)}\right)$, under the conditions in Theorem~\ref{Th1} and Condition~\ref{con: min-signal}, we have  
	\begin{align*}
	P\left(|\widehat{\mathcal{S}}_{v_n}|\leq  O(n^{\eta+\alpha-\omega/2})\right)\geq 1-p\left[2\exp\left(-c_2 r^2_n n^{1-4\alpha}\right)+2\exp\left(-c_3 r_n n^{1-2\alpha}\right)\right]
	\end{align*}
	for all $n$ sufficiently large. 
\end{theorem}

\noindent{\bf Remark 3}.  We would like to emphasize 
that the threshold $\nu_n$ and the set $\widehat{\mathcal{S}}_{v_n}$ in \eqref{eq: S-vn} are introduced primarily for the purpose of theoretical development.  We do not use the threshold $\nu_n$ when implementing our screening method (see AQFS in Algorithm 1).  In addition, 
Theorem~\ref{Th: size} suggests that the size of 
the set $\widehat{\mathcal{S}}_{v_n}$ is of polynomial order of $n$ with probability tending to one.  This result, together with Theorem~\ref{Th2}, indicates that if we choose the first $K_n$ covariates sequentially using our screening algorithm AQFS with $K_n=[ n^{\varsigma+\eta+\alpha-\omega/2}/\log (n)]$ for some $\varsigma>0$, then all relevant covariates can be retained with probability tending to one by noting that $K_n=[ n^{\varsigma+\eta+\alpha-\omega/2}/\log (n)]=[ n^{\eta+\alpha-\omega/2}\cdot n^{\varsigma}/\log (n)]\geq O(n^{\eta+\alpha-\omega/2})$. 
By assuming $\eta<1+\omega/2-\alpha$ and taking 
$\varsigma=1+\omega/2-\alpha-\eta$, we have $K_n=[ n/\log (n)]$. This is the reason why we suggest to take $K_n=[ n/\log (n)]$ in our AQFS algorithm. 
Such a choice of $K_n$ results in the estimated active set containing $K_n=[ n/\log (n)]$ covariates after the screening stage. 
The same choice for the number of covariates kept in the screening stage is 
also widely used in the variable screening literature; see, for example, \cite{fanlv2008}, \cite{fan2011},  \cite{li2012feature}, \cite{he2013},  \cite{maetal2017}, 
\cite{zhong2023feature}, and references therein.

As stated in Section~\ref{sec:method}, our AQFR algorithm has generated the $K_n$ nested models  $\mathcal{S}^{(1)}\subset \mathcal{S}^{(2)}\subset \cdots \subset \mathcal{S}^{(K_n)}$ and we can use QBIC  
to find the best model $\hat{\mathcal{S}}_{\mathrm{QBIC}}$ among those $K_n$ models after screening.
To establish the consistency of QBIC, let us introduce some additional notation. 
Recall that $\bz_{\mathcal{S}}=(1, \bPi^{\top}_{\mathcal{S}})^{\top}$ where $\bPi_{\mathcal{S}}$ is the column vector formed by stacking vectors $\bB_j(X_j)$, $j\in\mathcal{S}$, and $\bz_{i,\,\mathcal{S}}$ is the value of $\bz_{\mathcal{S}}$ at $\bx_i=(X_{i1}, \cdots, X_{ip})^{\top}$ from the $i$th observation. Similarly, let $\bz=(1, \bPi^{\top})^{\top}$ where $\bPi$ is the column vector formed by stacking vectors $\bB_j(X_j)$, $j\in\{1, \cdots, p\}$, and $\bz_{i}$ is the value of $\bz$ at $\bx_i$. For each $j=1, \cdots, p$, let $B_j^{\top}(X_{ij})\btheta_j^{*}$ be the best approximation, in the sup-norm, of $g_{j,\,\tau}(X_j)$ and $\Delta_j^{*}$ be the corresponding approximation error, i.e.,  $\Delta_j^{*}=\sup_{X_j}|B_j^{\top}(X_{j})\btheta_j^{*}-g_{j,\,\tau}(X_{j})|$. Write  $\Delta^{*}=\sup_{\bx}|\bz^{\top}\btheta^{*}-\mu_{\tau}-\sum_{j=1}^pg_{j,\,\tau}(X_{j})|$ where $\btheta^{*}=(\mu_{\tau}, {\btheta^{*}_1}^{\top}, \cdots, {\btheta^{*}_p}^{\top})^{\top}$.
Let $U=Y-\mu_{\tau}-\sum_{j=1}^pg_{j,\,\tau}(X_j)=Y-\mu_{\tau}-\sum\limits_{j\in\mathcal{S^*}}g_{j,\,\tau}(X_j)$ and $U_i=Y-\mu_{\tau}-\sum\limits_{j\in\mathcal{S^*}}g_{j,\,\tau}(X_{ij})$ for $i=1, \cdots, n$. We impose the following conditions to establish the screening consistency of QBIC.

\begin{condition}\label{con: error-U}
	The conditional distribution $F_{U|\bx}(\cdot|\bx)$ of the error $U$ given $\bx$ has a density $f_{U|\bx}$ such that (i) $\sup_{u, \bx}f_{U|\bx}(u|\bx)<\infty$ and (ii) there exist positive constants $\delta_1$ and $\delta_2$ satisfying  $\inf_{\bx}\inf_{|u|\leq \delta_1} f_{U|\bx}(u|\bx)\leq \delta_2$.
\end{condition}

\begin{condition}\label{con: eigenvalue}
	(i) $\max_{1\leq i\leq n}\max_{1\leq j\leq p}\|B_j(X_{ij})\|=O(\sqrt{q_n})$; (ii) $\Delta_j^{*}=O(q_n^{-d})$ for all $1\leq j\leq p$ and $\Delta^{*}=O(q_n^{-d})$; (iii) There exists positive finite constants $\lambda_1$ and $\lambda_2$ such that 
	\begin{align*}
	\lambda_1\leq \lambda_{\min}(E(\bz_{\mathcal{S}}\bz_{\mathcal{S}}^{\top}))\leq \lambda_{\max}(E(\bz_{\mathcal{S}}\bz_{\mathcal{S}}^{\top}))\leq \lambda_2
	\end{align*}
	uniformly for any subset $\mathcal{S}\subset\{1, 2, \cdots, p\}$ satisfying $|\mathcal{S}|=O(n^{\eta+\alpha-\omega/2})$, where $\lambda_{\min}(\cdot)$  and $\lambda_{\max}(\cdot)$ are the smallest and largest eigenvalues of a square matrix.
\end{condition}

Conditions~\ref{con: error-U} and~\ref{con: eigenvalue} correspond to Conditions (B1) and (B2) of \cite{LeeNohPark2014}, respectively. Condition~\ref{con: eigenvalue} is also used in~\cite{maetal2017}. 
In particular, as mentioned by \cite{LeeNohPark2014}, the first two requirements of Condition~\ref{con: eigenvalue} are satisfied for the spline basis approximation; see, for example, \cite{Horowitz2005nonparametric} and \cite{huang2010variable}.
Condition~\ref{con: eigenvalue} (iii) is the sparse Riesz condition, which places some restriction on the eigenvalues of 
$E(\bz_{\mathcal{S}}\bz_{\mathcal{S}}^{\top})$. This condition is commonly used in the literature to control the colinearity of covariates at the pouplation level in the high and ultra-high dimensional settings.

\begin{theorem}\label{Th4}
	Under all the conditions given in Theorem~\ref{Th: size} and Conditions~\ref{con: error-U} -\ref{con: eigenvalue}, and assuming that $\eta+\alpha-\omega/2<(2d-1)/(4d+2)$, 
	$C_n^{-1}=o(1)$, $n^{(\eta+\alpha-\omega/2)-1}C_nq_n\log(n)=o(1)$, and $E(|U|)<\infty$, we have
	$P(\mathcal{S}^{*} \subset  \hat{\mathcal{S}}_{\mathrm{QBIC}})  \rightarrow 1$ 	
	as $n\to\infty$. 
	
\end{theorem}

Theorem~\ref{Th4} shows that the best model selected by QBIC after the screening step, $\widehat{\mathcal{S}}_{\mathrm{QBIC}}$, also enjoys the sure screening property. That is, all the relevant covariates are contained in the selected model $\widehat{\mathcal{S}}_{\mathrm{QBIC}}$ with probability tending to one as $n\to\infty$.


\section{Simulation Studies \label{sec:simulation}}

In this section, we examine the numerical performance of the proposed screening  ($\mathcal{S}^{(K_n)}$) and selection ($\widehat{\mathcal{S}}_{\mathrm{QBIC}}$) using simulated data. 
For ease of reference, we will use the labels 
(AQFS) and (AQFS + QBIC) in place of 
$\mathcal{S}^{(K_n)}$ and  $\widehat{\mathcal{S}}_{\mathrm{QBIC}}$, respectively. To offer practical guidance on the choice of $C_n$, we consider three versions of AQFS + QBIC corresponding to different choices of $C_n$, see (\ref{eq: QBIC}).  When using $C_n=\log(\log p)$, we name the resulting selected set AQFS + QBIC1. To reduce possible under-fitting by the first choice, we consider $C_n=\log(\log p^{0.75})$ which is named AQFS + QBIC2. The third is even less stringent $C_n=\log(\log p^{0.5})$ leading to AQFS + QBIC3. 
We use $(n, p)=(300, 3000)$ in all simulation studies. 
We take $q_n=[n^{1/5}]=3$ by following~\cite{he2013}.

We repeat each simulation 100 times, and evaluate the performance through the following measures: 
{\it Screening of $X_j$}, the proportion that the important covariate $X_j$ is identified in the screening stage; {\it Sure screening rate} (All), the proportion that all important covariates are retained in the screening or selection stage; {\it Selection of $X_j$}, the proportion that the important covariate $X_j$ is identified in the selection stage; {\it False Positives} (FP), the average number of unimportant covariates incorrectly included in the selection stage; {\it False Negatives }(FN), the average number of important covariates missed in a selection stage; {\it Quantile prediction error} (QPE), based on an independent 5000 test observations (generated randomly from the true model), the QPE at some quantile $\tau$ is computed as $5000^{-1} \sum^{5000}_{i=1}(Y_i - \hat{Y}_i(\tau))$, where $\hat{Y}_i(\tau)$ is the quantile estimate of fitting an additive quantile regression based on selected covariates.

We consider three model settings that are special cases of the additive model~\eqref{eq: AQR-model}.  
As an added feature, we  introduce a heterogeneous covariate $X_1$ in the sense of having a varying association with the response $Y$ at different quantiles. As such, the ability to correctly identify such a heterogeneous covariate in the presence of ultra-high dimensional data is an important performance measure, among others.     

\medskip 
\noindent
{\bf Example 1} (fully linear). 
We consider a fully linear additive model by following~\cite{wangwuli2012} 
$$Y = X_6 + X_{12} + X_{15} + X_{20} + 0.7 X_1 \varepsilon,$$ 
where $\varepsilon$ is generated from the standard normal distribution $N(0, 1)$ and the covariates $X_1, \cdots, X_p$ are defined as follows. 
First, generate $\tilde{X}_1, \ldots, \tilde{X}_p$ from the multivariate normal distribution $N(\bzero, \bSig)$ where $\bSig=(\sigma_{jk})_{p\times p}$ with $\sigma_{jk} = 0.5^{|j-k|}$.
Then, define $X_1 = \sqrt{12} \, \Phi(\tilde{X}_1)$, where $\Phi$ is the cumulative distribution function of $N(0, 1)$ and $X_j = \tilde{X}_j$ for $j\geq 2$. As in \cite{wangwuli2012}, we take $\tau =0.3, 0.5$ and $0.7$.  The index set of truly relevant covariates $\mathcal{S}^*_\tau$, defined in \eqref{eq:true}, is 
$\{1, 6, 12, 15, 20\}$ 
at $\tau\neq 0.5$ while $\mathcal{S}^*_\tau  = \{6, 12, 15, 20\}$ at $\tau=0.5$.

\begin{table}[h]
	\centering
	\caption{Screening performance of different methods for Example 1 with $(n, p)=(300, 3000)$.	
	} \label{tab:screen1}
	\vspace{0.1in}
	\begin{tabular}{llcccccc}
		\hline
		$\tau$     & Method     & $X_1$   & $X_6$   & $X_{12}$  & $X_{15}$  & $X_{20}$  & All \\
		\hline
		0.3 & QSIS       & 0.06 & 1.00 & 1.00 & 1.00 & 1.00 & 0.06 \\
		& QaSIS      & 0.04 & 1.00 & 1.00 & 1.00 & 1.00 & 0.04 \\
		& AQFS       & 0.82 & 1.00 & 1.00 & 1.00 & 1.00 & 0.82 \\ 
		\hline
		0.5 & QSIS   &  --    & 1.00 & 1.00 & 1.00 & 1.00 & 1.00 \\
		& QaSIS      &  --    & 1.00 & 1.00 & 1.00 & 1.00 & 1.00 \\
		& AQFS       &  --    & 1.00 & 1.00 & 1.00 & 1.00 & 1.00 \\ 
		\hline
		0.7 & QSIS       & 0.16 & 1.00 & 1.00 & 1.00 & 0.99 & 0.16 \\
		& QaSIS      & 0.10 & 1.00 & 1.00 & 1.00 & 1.00 & 0.10 \\
		& AQFS       & 0.80 & 1.00 & 1.00 & 1.00 & 1.00 & 0.80 \\
		\hline
	\end{tabular}	
\end{table}

The variable screening results for Example 1 are summarized in Table \ref{tab:screen1}. We also include two existing non-parametric screening methods, QSIS~\citep{wuyin2015} and QaSIS~\citep{he2013}, in our comparison. 
To ensure comparability, the number of retained covariates is $K_n=[n/\log(n)]=52$ for all screening methods. 
As expected, QSIS, QaSIS, and AQFS all can identify important covariates $X_{6}$, $X_{12}$, $X_{15}$ and $X_{20}$. However, 
as reflected in the low screening rates at $\tau=0.3$ and $0.7$, both QSIS and QaSIS have difficulty in identifying $X_1$. In contrast, our method AQFS detects $X_1$ at significantly improved rates. Such a competitive performance  highlights the value of the proposed screening algorithm AQFS.

\begin{table}[h]
		\centering
	\caption{Selection performance of different methods for Example 1 with $(n, p)=(300, 3000)$. For the quantile prediction error (QPE), values inside the parentheses are  
    standard errors.	
	The QPE values for AQFS + GamBoost are not shown because the estimation of its additive components cannot be directly comparable to those of AQFS + QBIC.}		
	\label{tab:select1}
	\vspace{0.1in}
	\scalebox{1}{
		\begin{tabular}{llccccccccc}
			\hline
			$\tau$     & Method          & $X_1$   & $X_6$   & $X_{12}$   & $X_{15}$   & $X_{20}$   & All  & FP    & FN   & QPE          \\
			\hline
			0.3 & Q-SCAD      & 0.15 & 1.00	& 1.00 & 1.00 &	1.00 & 0.15	& 294.24 & 0.85	& 0.71 (0.00) \\
			    & AQFS+QBIC1      & 0.58 & 1.00 & 1.00 & 1.00 & 1.00 & 0.58 & 0.19  & 0.42 & 0.46 (0.00) \\
			    & AQFS+QBIC2      & 0.64 & 1.00 & 1.00 & 1.00 & 1.00 & 0.64 & 0.47  & 0.36 & 0.46 (0.00) \\
			    & AQFS+QBIC3      & 0.73 & 1.00 & 1.00 & 1.00 & 1.00 & 0.73 & 1.08  & 0.27 & 0.47 (0.00) \\
			    & AQFS+GamBoost   & 0.01 & 1.00 & 1.00 & 1.00 & 1.00 & 0.01 & 0.00  & 0.99 & -- \\ 
			    & Oracle          &  --    &  --    &   --   &   --   &  --    &   --   &  --     &   --   & 0.45 (0.00) \\ 
			\hline
			0.5 & Q-SCAD      &	 --    & 1.00	& 1.00 & 1.00 &	1.00 & 1.00	& 294.97 &0.00 & 0.71 (0.00) \\
			    & AQFS+QBIC1  &  --    & 1.00 & 1.00 & 1.00 & 1.00 & 1.00 & 0.02  & 0.00 & 0.51 (0.00) \\
			& AQFS+QBIC2      & --     & 1.00 & 1.00 & 1.00 & 1.00 & 1.00 & 0.05  & 0.00 & 0.51 (0.00) \\
			& AQFS+QBIC3      &  --    & 1.00 & 1.00 & 1.00 & 1.00 & 1.00 & 0.29  & 0.00 & 0.51 (0.00) \\
			& AQFS+GamBoost   &  --    & 1.00 & 1.00 & 1.00 & 1.00 & 1.00 & 0.00  & 0.00 & -- \\ 
			& Oracle          & --     &  --    &   --   &  --    &  --    &  --    &  --     &  --    & 0.51 (0.00) \\
			\hline
			0.7 & Q-SCAD      &	0.12 & 1.00	& 1.00 & 1.00 &	1.00 & 0.12	& 294.04 & 0.88	& 0.71 (0.00) \\
			& AQFS+QBIC1      & 0.51 & 1.00 & 1.00 & 1.00 & 1.00 & 0.51 & 0.21  & 0.49 & 0.47 (0.00) \\
			& AQFS+QBIC2      & 0.62 & 1.00 & 1.00 & 1.00 & 1.00 & 0.62 & 0.52  & 0.38 & 0.47 (0.00) \\
			& AQFS+QBIC3      & 0.70 & 1.00 & 1.00 & 1.00 & 1.00 & 0.70 & 1.02  & 0.30 & 0.47 (0.00) \\
			& AQFS+GamBoost   & 0.00 & 1.00 & 1.00 & 1.00 & 1.00 & 0.00 & 0.01  & 1.00 & -- \\
			& Oracle          &   --   &  --    &   --   &  --    &  --    & --     &  --     & --     & 0.45 (0.00) \\ 
			\hline      
	\end{tabular}}
\end{table}

The variable selection results for Example 1 are summarized in Table \ref{tab:select1}. 
We include the 
SCAD-penalized regression method (abbreviated as Q-SCAD) for linear quantile regression model in~\cite{wangwuli2012} in our comparison.
With the goal to serve as a benchmark for prediction performance, we also include 
the oracle approach, 
which assumes that the true set of important covariates is known.
With the exception for $X_1$, the retention rate of the important covariates by AQFS is fully preserved by the selection stage (for all the three versions of AQFS + QBIC). The least stringent penalty, i.e. AQFS + QBIC3, performs best in terms of retaining $X_1$ without compromising on prediction accuracy (QPE). 
When comparing Q-SACD with our methods, we can observe the following from  Table~\ref{tab:select1}: (i) Both Q-SCAD and our methods can select all important covariates at $\tau=0.5$; (ii) Compared to our methods, Q-SCAD has poor performance in identifying $X_1$ at $\tau=0.3$ and $\tau=0.7$; (iii) Q-SCAD is much worse than our methods in terms of false positives (FP); (iv) Our methods have smaller values for false negatives (FN) and quantile prediction error (QPE). Overall, our methods exhibit the better performance than Q-SCAD in this example. Note that Q-SCAD knows the linearity of additive terms but all versions of AQFS+QBIC do not. This further highlights the value of the proposed method.

Recall that all the selection approaches AQFS + QBIC are based on the idea of picking the best model from 
the
$K_n$ nested models $\mathcal{S}^{(1)}\subset \mathcal{S}^{(2)}\subset \cdots \subset \mathcal{S}^{(K_n)}$.
As a possible alternative, one may also consider the largest screened set from AQFS, i.e. $\mathcal{S}^{(K_n)}$, and apply shrinkage or penalized additive quantile regression on this set. 
One such shrinkage approach is the boosting algorithm in \cite{Fenskeetal2011GAMBOOST}, called GamBoost.
We employ the R package~\texttt{mboost} to implement GamBoost and tune the parameter $m_{\mathrm{stop}}$ (which is the number of boosting iterations in GamBoost) by 10-fold cross-validation. As shown in Table~\ref{tab:select1},
AQFS + GamBoost, performs similarly to those based on QBIC with the exception of $X_1$. However, AQFS + GamBoost appears to lack power to identify the heterogeneous covariate at $\tau=0.3$ and $0.7$.


\bigskip \noindent
{\bf Example 2} (Mostly linear).  
This example is adapted from \cite{sherwood2016}. We consider the model
\begin{align*}
Y = \beta_1 X_6 +  \beta_2 X_{12} + \beta_3 X_{15} + \beta_4 X_{20} + g_{25}(X_{25}) + g_{26}(X_{26}) + 0.7 X_1 \varepsilon,
\end{align*}
where $\beta_j \sim U[0.5, 1.5]$ for $1 \leq j \leq 4$,
$g_{25}(X_{25})=\sin(2 \pi X_{25})$, $g_{26}(X_26)=2.5 X^3_{26}$
and $\varepsilon$ is generated from the standard normal distribution $N(0, 1)$. The covariates $X_1, \ldots, X_p$ are defined as follows. 
First, generate $\tilde{X}_1, \ldots, \tilde{X}_p$ from the multivariate normal distribution $N(\bzero, \bSig)$ where $\bSig=(\sigma_{jk})_{p\times p}$ with $\sigma_{jk} = 0.5^{|j-k|}$.
Then, define $X_1 = \sqrt{12} \Phi(\tilde{X}_1)$, $X_{25} =  \Phi(\tilde{X}_{25})$ and $X_{26} =  \Phi(\tilde{X}_{26})$ where $\Phi(\cdot)$ is the cumulative distribution function of $N(0, 1)$ and $X_j = \tilde{X}_j$ for the rest of covariates. 
As in \cite{sherwood2016}, we consider $\tau =0.5$ and $0.7$. 
It is seen that the index set of truly relevant covariates  $\mathcal{S}^*_\tau  = \{6, 12, 15, 20, 25, 26\}$ at $\tau=0.5$ while $\mathcal{S}^*_\tau=\{1, 6, 12, 15, 20, 25, 26\}$ at $\tau=0.7$.

The variable screening results for Example 2 are summarized in Table \ref{tab:screen2}.
It is seen that QSIS, QaSIS, and AQFS all can identify important linear covariates $X_{6}$, $X_{12}$, $X_{15}$ and $X_{20}$, but QSIS and QaSIS have difficulty in identify important nonlinear covariates $X_{25}$ and $X_{26}$.
AQFS significantly outperforms both QSIS and QaSIS, at both quantiles $\tau=0.5$ and $0.7$. This is not surprising because both QSIS and QaSIS use marginal screening and can miss important covariates. 
In addition, AQFS also perform better than QSIS and QaSIS in terms of detecting the heterogeneous covariate $X_1$ at $\tau=0.7$.

\vspace{-2mm}
\begin{table}[H]
	\centering
	\caption{Screening performance of different methods for Example 2 with $(n, p)=(300, 3000)$.
	} \label{tab:screen2} 
	\vspace{0.1in}
		\scalebox{0.9}{
	\begin{tabular}{llcccccccc}
		\hline
		$\tau$      & Method           & $X_1$   & $X_6$                & $X_{12}$  & $X_{15}$  & $X_{20}$  & $X_{25}$  & $X_{26}$  & All  \\
		\hline
		0.5 & QSIS             &   --   & 0.94              & 0.99 & 1.00 & 1.00 & 0.24 & 0.58 & 0.05 \\
		& QaSIS            &   --   & 0.94              & 0.99 & 1.00 & 1.00 & 0.42 & 0.45 & 0.17 \\
		& AQFS             & --    & 1.00              & 1.00 & 1.00 & 1.00 & 0.95 & 0.98 & 0.95 \\ 
		\hline
		0.7 & QSIS             & 0.15 & 0.90              & 0.98 & 1.00 & 1.00 & 0.26 & 0.46 & 0.00 \\
		& QaSIS            & 0.04 & 0.82              & 0.97 & 1.00 & 1.00 & 0.47 & 0.29 & 0.00 \\
		& AQFS             & 0.58 & 0.99              & 1.00 & 1.00 & 1.00 & 0.96 & 0.96 & 0.55 \\
		\hline
	\end{tabular}}	
	
\end{table}

The variable selection results for Example 2
are summarized in Table~\ref{tab:select2}.
We include the SCAD-penalized regression method incorporating non-linear terms (abbreviated as QN-SCAD) in~\cite{sherwood2016} in our comparison.
Since QN-SCAD assumes that the model structure is known and focuses on selecting important covariates
in the linear part only, no selection rates for $X_{25}$ and $X_{26}$ are reported for QN-SCAD. 
We observe the following from Table~\ref{tab:select2}: (i) All methods work really well in terms of selecting all important covariates
in the linear part at $\tau=0.5$ (i.e., $X_6$, $X_{12}$, $X_{15}$ and  $X_{20}$ in this case); (ii) QN-SCAD is much worse than our methods in terms of false positives (FP); (iii) Our methods have smaller value for QPE. Overall, our methods exhibit the better performance than QN-SCAD.

\begin{table}[H]
	\caption{Selection performance of different methods for Example 2 with $(n, p)=(300, 3000)$. For the quantile prediction error (QPE), values inside the parentheses are  
	standard errors.}
	\label{tab:select2}
	\vspace{0.1in}	
	\scalebox{0.90}{
		\begin{tabular}{llccccccccccc}
			\hline
			$\tau$      & Method                 & $X_1$  & $X_{6}$   & $X_{12}$  & $X_{15}$  & $X_{20}$  & $X_{25}$  & $X_{26}$  & All  & FP   & FN   & QPE          \\
			\hline
			0.5       & QN-SCAD       &   --   & 1.00 & 1.00	& 1.00 & 1.00 &	--	 &  --    &1.00 & 288.90 & 0.00	& 0.75 (0.01)\\
			          & AQFS+QBIC1    &  --    & 0.99 & 0.99 & 1.00 & 1.00 & 0.91 & 0.88 & 0.84 & 0.71  & 0.23 & 0.63 (0.01) \\
			& AQFS+QBIC2                &   --   & 0.99 & 0.99 & 1.00 & 1.00 & 0.92 & 0.90 & 0.86 & 0.83  & 0.20 & 0.63 (0.01) \\
			& AQFS+QBIC3                & --   & 1.00 & 0.99 & 1.00 & 1.00 & 0.93 & 0.94 & 0.89 & 1.13  & 0.14 & 0.63 (0.01) \\
			& Oracle                    &  --    & --     &   --   &  --    &   --   &   --   &  --    & --     &    --   &  --    & 0.61 (0.01) \\
			\hline
			0.7       & QN-SCAD       &   0.19	& 1.00	& 1.00	& 1.00	& 1.00	& -- & -- &	0.19 &	287.57 &	0.81 &	0.71 (0.01)\\
		               & AQFS+QBIC1                & 0.19 & 0.93 & 0.99 & 1.00 & 1.00 & 0.76 & 0.80 & 0.19 & 0.95  & 1.33 & 0.57 (0.01) \\
			& AQFS+QBIC2                & 0.24 & 0.97 & 0.99 & 1.00 & 1.00 & 0.87 & 0.87 & 0.24 & 1.45  & 1.06 & 0.56 (0.01) \\
			& AQFS+QBIC3                & 0.34 & 0.97 & 0.99 & 1.00 & 1.00 & 0.90 & 0.90 & 0.32 & 1.87  & 0.90 & 0.57 (0.01) \\
			& Oracle                    &   --   &   --   &  --    &   --   &  --    &  --    &   --   &  --    &  --     &  --    & 0.53 (0.00) \\
			\hline
		\end{tabular}
	}
\end{table}


\medskip 
\noindent
{\bf Example 3} (mostly nonlinear). 
This example is adapted from \cite{LeeNohPark2014}. We consider the model
$$
Y = 5 X_2 + g_3(X_3) + g_4(X_4) + g_1(X_1) \varepsilon,
$$
where $\varepsilon$ is generated from the standard normal distribution $N(0, 1)$. 
Each covariate $X_j$ is defined by
$X_j = 0.5(\tilde{X}_j + U)$ for $j=1, \dots, p$ where $\tilde{X}_1, \dots, \tilde{X}_p$ and $U$ are independently generated from the standard Uniform distribution.  
It can be seen that the covariates $X_1, \cdots, X_p$ are highly correlated because the pairwise population correlation between $X_j$ and $X_k$ is 0.5 for $j\ne k$.  
The nonlinear additive functions are given by $g_1(X_1) = 7X^2_1$, ~$g_3(X_3) =  4\sin(2\pi X_3) / ( 2-\sin(2\pi X_3) )$ and 
$g_4(X_4) = 0.6\sin(2\pi X_4) + 1.2\cos(2\pi X_4) + 1.8\sin^2(2\pi X_4) + 2.4\cos^3(2\pi X_4) + 3 \sin^3(2\pi X_4)$. 
As in  \cite{LeeNohPark2014}, we consider $\tau =0.2$ and $0.5$.  
We can see that the index set of truly important covariate $\mathcal{S}^*_\tau= \{ 1, 2, 3, 4\}$ at $\tau\neq 0.5$ while $\mathcal{S}^*_\tau  = \{2, 3, 4 \}$ at $\tau=0.5$.  
Note that the heterogeneous covariate $X_1$ also has a nonlinear association with the response $Y$ at quantiles
other than $\tau=0.5$.

The variable screening results for Example 3 are summarized in Table \ref{tab:screen3}.  
We can see that both QSIS and QaSIS have no power in identifying $X_2$ at quantiles $\tau=0.2$ and $0.5$.
This is because QSIS and QaSIS use marginal screening and can miss important covariates when the variables are highly correlated. In addition, compared to QaSIS and AQFS, QSIS has lower power to identity nonlinear covariate $X_4$. This is not surprising because QSIS is designed for linear quantile regression.
It is interesting to observe that both QSIS and QaSIS
is comparatively competitive with AQFS in identifying in the heterogeneous covariate $X_1$ in this Example. 
This may be due to its high correlations with other important predictors $X_2$, $X_3$ and $X_4$.

The variable selection results for Example 3 
are summarized in Table \ref{tab:select3}.
Overall, the retention rate of important covariates is largely preserved by the selection stage, especially by AQFS + QBIC3.

\begin{table}[H]
	\centering
	\caption{Screening performance of different methods for Example 3 with $(n, p)=(300, 3000)$.} 
	\label{tab:screen3}
	\vspace{0.1in}
	\begin{tabular}{llccccc}
		\hline
		$\tau$   & Method    & $X_1$   & $X_2$   & $X_3$   & $X_4$   & All  \\
		\hline
		0.2      & QSIS      & 0.50 & 0.00 & 0.99 & 0.55 & 0.00 \\
		         & QaSIS     & 0.59 & 0.00 & 0.89 & 0.95 & 0.00 \\
		         & AQFS      & 0.63 & 0.88 & 0.95 & 0.93 & 0.59 \\ 
		\hline
		0.5 & QSIS  & -- & 0.00 & 1.00 &   1.00   & 0.00 \\
		& QaSIS     & -- & 0.00 & 0.98 &   1.00   & 0.00 \\
		& AQFS      & -- & 1.00 & 1.00 &   1.00   & 1.00 \\
		\hline
	\end{tabular}
	
\end{table}

\begin{table}[H]
	\centering
	\caption{Selection performance of different methods for Example 3 with $(n, p)=(300, 3000)$. For the quantile prediction error (QPE), values inside the parentheses are 
	standard errors.} \label{tab:select3}
	\vspace{0.1in}	
	\begin{tabular}{llcccccccc}
		\hline
		$\tau$     & Method                    &$X_1$   & $X_2$   & $X_3$   & $X_4$   & All  & FP    & FN   & QPE          \\
		\hline
		0.2 & AQFS+QBIC1  & 0.41 & 0.64 & 0.87 & 0.92 & 0.37 & 0.72  & 1.16 & 0.69 (0.01) \\
	  	    & AQFS+QBIC2  & 0.53 & 0.76 & 0.91 & 0.93 & 0.48 & 1.52  & 0.87 & 0.69 (0.01) \\
		    & AQFS+QBIC3  & 0.56 & 0.83 & 0.93 & 0.93 & 0.52 & 2.79  & 0.75 & 0.72 (0.01) \\
		    & Oracle      &   -- & --   &  --  &  --  &  --  & --    &  --  & 0.62 (0.00) \\ 
		\hline
		0.5 & AQFS+QBIC1  & -- & 0.98 & 1.00 &   1.00   & 0.98 & 0.04  & 0.02 & 0.86 (0.00) \\
		    & AQFS+QBIC2  & -- & 0.98 & 1.00 &  1.00    & 0.98 & 0.10  & 0.02 & 0.86 (0.00) \\
		    & AQFS+QBIC3  & -- & 0.98 & 1.00 &   1.00   & 0.98 & 0.26  & 0.02 & 0.87 (0.00) \\
		    & Oracle      & -- &  --  & --   &  --    &  --    & --      &  --    & 0.86 (0.00) \\
		\hline
	\end{tabular}
	
\end{table}

In summary, through three examples above, our proposed variable screening algorithm, AQFS, is shown to be powerful in identifying important covariates. For a dimension $p$ as large as $3000$, the screening method is also reasonably effective in detecting a covariate that has heterogeneous association with the response. The covariate selection results are also encouraging and the least stringent penalty with the lowest $C_n$, i.e. AQFS + QBIC3, performs the best overall. In terms of prediction, the simulation results show that the proposed selection competes even with 
the oracle method which uses the information of the true underlying sparse model (i.e., the model structure and which covariates are important). Finally, our procedure is computationally fast to be implemented. 
This is very appealing for practitioners when faced with ultra-high dimensional applications.

\section{Real Data Example \label{sec:application} }

We consider the problem of identifying potential risk factors for low infant {\it birth weight}. Following \cite{sherwood2016}, we use data on 64 subjects to select important risk factors at quantiles in lower half of the conditional distribution of infant birth weight (in kilograms).  We focus on the quantiles $\tau=0.2$, $0.3$ and $0.5$. 
The covariate set (potential risk factors) has a total dimension of $p=24531$, i.e. 5 clinical covariates and expression levels of 24526 different probes. The genetic data are from the peripheral blood sample collected by \cite{votavova2011}. The clinical covariates are {\it age} of the mother, {\it gestational age}, {\it parity}, the amount of {\it continine} in the blood (which is a chemical found in tobacco) and mother's {\it BMI}. Although the proposed AQFS is still feasible for such large dimension $p$, it may lack power to detect important covariates given a small sample of $n=64$. To help explore this issue, we consider four covariate dimensions $p=m+5$ where 
$m\in\{24526, 1200, 600, 200\}$. 
In other words, 
$p\in\{24531, 1205, 605, 205\}$.
The candidate set of $m$ probes is obtained as follows. At step $\ell=2$ in our screening algorithm AQFS (see step $2(b)$ of our Algorithm~\ref{algorithm:AQFR}), instead of selecting one probe that achieves the maximum score, we pick the top $m$ probes. Then we can apply 
our procedure to the set with $m+5$ covariates (i.e., $m$ probes and $5$ clinical covariates) for variable screening. 

In order to evaluate the performance of our procedure, we consider 100 random partitions of the original data into a training set of 58 (new $n$) observations and the rest 6 as test data. We opt for this 90\%-10\% split to conserve data for training the AQFS algorithm 
as the initial $n=64$ is rather small. For all considered $p$, the top 14 ($=[58/ \log(58)]$) covariates are 
kept in the screening stage. To possibly further reduce these screened sets, we apply GamBoost in \cite{Fenskeetal2011GAMBOOST} for further variable selection.  
We tune the parameter $m_{\mathrm{stop}}$ in GamBoost via 10-fold cross-validation.  
Based on the 100 random partitions, Table ~\ref{table:realdata} reports the average quantile prediction error (QPE) evaluated on the test data and the average model size (MS). We also list the most frequently selected covariates (with at least 15\% relative frequency) for each $p$ and $\tau$.

For all $\tau$'s, 
our results across $p$ are comparable both in terms of QPE and MS.
As $p$ increases, a smaller number of covariates are frequently selected, indicating the difficulty of separating noises from true signals given the small sample. Interestingly, those covariates selected at least 40\% of times in $p=205$ and $p=605$ are also selected with high frequency in large $p$ cases. So, when the signal is large enough, AQFS can detect important covariates even in small samples from the ultra-high dimensional set. For all $\tau$'s, the two clinical covariates, {\it Age}  and {\it continine} level, are by far the most frequently selected. For the full dimension case $p=24531$, {\it continine} is not as frequently selected as in the smaller $p$ cases. Among the other  clinical covariates, {\it gestational age} is also frequently selected at $\tau=0.2$ and {\it parity} at all three $\tau$'s.

\begin{figure}[h!]
	\centering
	\caption{\label{fig:additiveC} Plots of the estimated effects of the  selected clinical variables at different quantiles (row 1: $\tau=0.2$; row 2: $\tau=0.3$; row 3: $\tau=0.5$). }		
	\includegraphics[width=17cm, height=15cm]{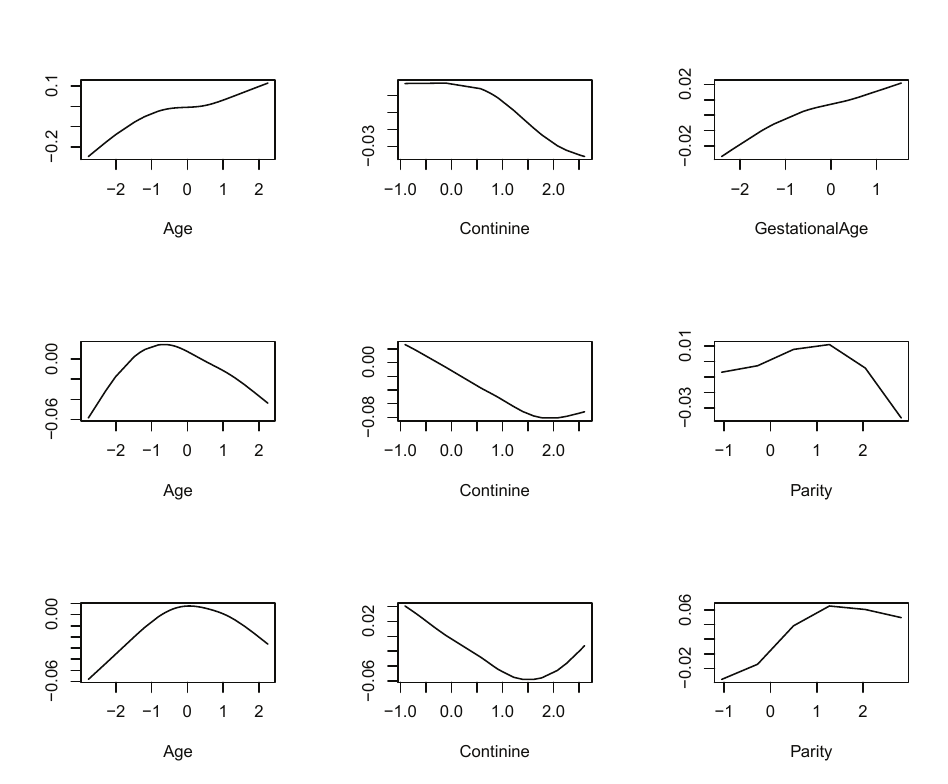}
\end{figure}

\begin{figure}[h!]
	\centering
	\caption{\label{fig:additiveP} 
		Plots of the estimated effects of the top three selected probes at different quantiles (row 1: $\tau=0.2$; row 2: $\tau=0.3$; row 3: $\tau=0.5$). }	
	\includegraphics[width=17cm, height=15cm]{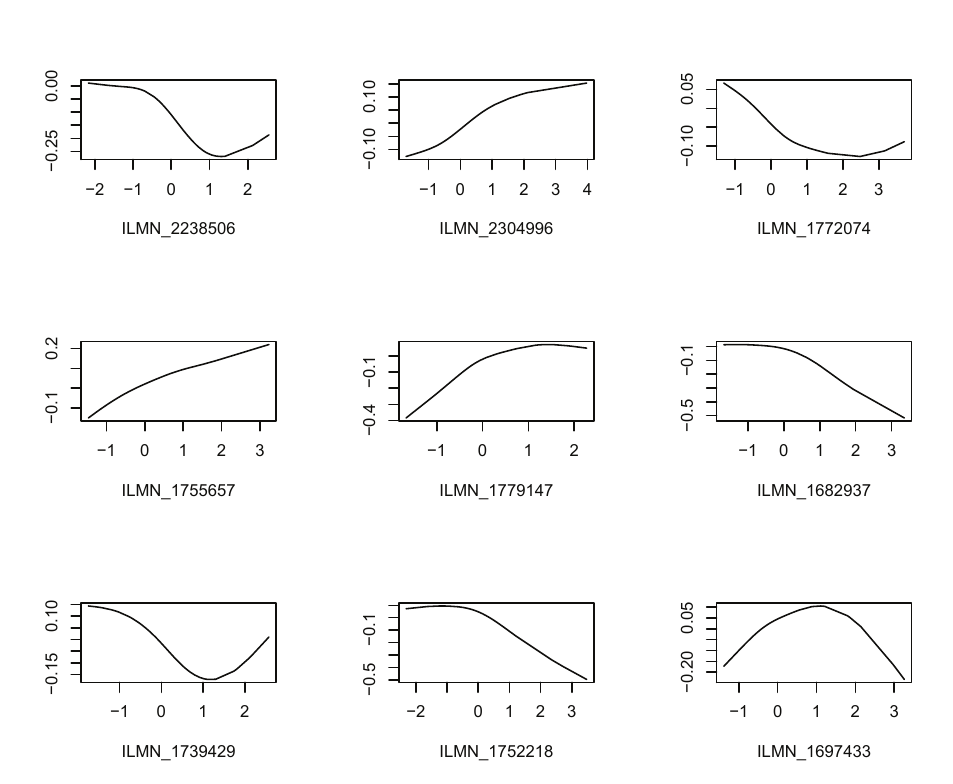}
\end{figure}

To explore the nature of the relationship between {\it birth weight} and clinical covariates for a particular $\tau$, we combine the most frequently selected variables (for all $p$) from Table~\ref{table:realdata}.  
This way, we have a total 14 unique covariates at $\tau=0.2$, 12 covariates at $\tau=0.3$, and 12 covariates at $\tau=0.5$, respectively.
Then, for each $\tau$, we apply GamBoost (further tuned with 10-fold cross-validation) on the original $n=64$.  
In particular, three clinical covariates, {\it age}, {\it continine} and {\it gestational age}, are selected at $\tau=0.2$. The covariate {\it parity} was selected in addition to {\it age} and {\it continine} at $\tau=0.3$ and $0.5$. 
Figure~\ref{fig:additiveC} shows the estimated 
effects (i.e., estimated functions $\hat{g}_j(X_j)$) of the selected clinical covariates at each $\tau$. 
We can observe that most of the estimated effects of those selected clinical covariates
are fairly nonlinear and not necessarily the same across quantiles.
We also display the estimated effects of three most frequently selected probes for each quantile $\tau$ in Figure~\ref{fig:additiveP}.
The nonlinear effects of those probes appear even stronger.

\cite{sherwood2016} analyzed the same data using a semiparametric quantile regression model at $\tau=0.1$, $0.3$ and $0.5$ and reported covariate selection results. To be more specific, for each quantile,  \cite{sherwood2016} first applied QaSIS of \cite{he2013} on the 24526 probes to select the top 200 probes. Then they included the gene expression values of
the 200 probes and the clinical variables {\it parity}, {\it gestational age}, {\it cotinine} and
{\it BMI} as linear covariates and the clinical variable {\it age} of the mother as nonlinear covariate. Their approach applies penalization (covariate selection) only on those linear covariates ($p=204$). They reported that, among the clinical covariates that are subject to penalization, only {\it gestational age} is selected to have important association with {\it birth weight}. Among the frequently selected probes, none of them match our selection except probe ILMN$\_$1755657. Overall, possibly due to the small sample, there is little overlap between the covariate selection results of AQFS and those of \cite{sherwood2016}. As such the findings from both approaches can be considered complementary.

\begin{table}[H]
	\centering
	\caption{\label{table:realdata} Selected covariates (relative frequency in parenthesis),  average quantile prediction error (QPE) and average model size (MS) over 100 random partitions. For PQE and MS, the values given in parenthesis are their
	standard errors.
	}
	\vspace{0.1in}
	\label{screenEx1}
	\scalebox{0.72}{	
		\begin{tabular}{lllll lllll}		
			\hline 
			& & $p=205$                & \textbf{}  & $p=605$             & \textbf{}  & $p=1205$             & \textbf{} & $p=24531$   \\  \hline 
			\multirow{13}{*}{$\tau=0.2$} & 	\multirow{11}{*}{Covariate}	  	     & Age (86\%)       &           & Age (79\%)      &           & Age (86\%)           &     &  Age (90\%)                &  \\
			& & Continine (62\%)&           & Continine (47\%)  &           &  Continine (47\%)     &     & ILMN$\_$2125395 (48\%)            &  \\
			& & ILMN$\_$1697444 (36\%)        &           & ILMN$\_$2159694 (28\%)         &           & ILMN$\_$2159694 (26\%)       &           & ILMN$\_$2159694 (33\% )           &  \\
			& & ILMN$\_$1785960 (33\%)        &           & ILMN$\_$1697444 (24\%)         &           & ILMN$\_$2355486 (22\%)       &           & ILMN$\_$2355486 (17\%)            &  \\
			& & ILMN$\_$2159694 (31\%)        &           & ILMN$\_$2355486 (21\%)         &           & ILMN$\_$1772074 (19\%)       &           & {\it Continine} (10\%)    &  \\
			& & ILMN$\_$2355486 (29\%)        &           & ILMN$\_$1772074 (19\%)   &           &                      &           &                           &  \\
			& & Parity (24\%)         &           & ILMN$\_$1785960 (17\%)         &           &                      &           &                           &  \\
			& & ILMN$\_$2238506 (21\%)        &           & ILMN$\_$1755657 (16\%)         &           &                      &           &                           &  \\
			& & GestationalAge (16\%) &           & ILMN$\_$2304996 (16\%)         &           &                      &           &                           &  \\
			& & ILMN$\_$1716061 (16\%)        &           &                        &           &                      &           &                           &  \\
			& & ILMN$\_$1772074 (16\%)        &           &                  &           &                      &           &                           & \\ 
			\cline{2-10}
			& QPE	& 0.141 (0.051)         &           &  0.135 (0.047)      &               & 0.133 (0.047)        &           & 0.147 (0.046)                          & \\
			& MS		    & 10.17 (2.01)          &           &  10.43 (2.19)       &               & 10.16 (1.94)         &           & 10.93 (1.89)                          & \\		\hline
			\multirow{13}{*}{$\tau=0.3$} & 	\multirow{11}{*}{Covariate}		        & Age (76\%)           &      & Age (91\%)          &        & Age (81\%)           &           &  Age (93\%)                &  \\
			& & Continine (71\%)      &     &  Continine (60\%)    &        & ILMN$\_$1800059 (53\%)     &           & ILMN$\_$1800059 (53\%)            &  \\
			& & ILMN$\_$1800059 (52\%)        &           & ILMN$\_$1800059 (53\%)      &              & Continine (45\%)       &           & ILMN$\_$2125395  (36\% )           &  \\
			& & ILMN$\_$2051113 (19\%)        &           & ILMN$\_$1772074 (26\%)      &        & ILMN$\_$2229649(28\%)       &           &  {\it Continine} (9\%)            &  \\
			& & ILMN$\_$1755657 (17\%)        &           & ILMN$\_$2229649(26\%)      &               &  ILMN$\_$1772074 (22\%)       &           &                            &  \\
			& & ILMN$\_$1682937 (16\%)        &           & ILMN$\_$1680434(17\%)      &           &                      &           &                           &  \\
			& & ILMN$\_$1772074 (16\%)         &           & Parity (16\%)      &           &                      &           &                           &  \\
			& & ILMN$\_$1779147 (16\%)        &           &                    &           &                      &           &                           &  \\ 
			\cline{2-10}
			& QPE    & 0.191 (0.055)         &           &  0.194 (0.056)      &           & 0.192 (0.072)        &           & 0.191 (0.065)                          & \\
			& MS 		    & 10.72 (1.81)          &           &  11.60 (1.67)       &           & 11.02 (1.84)         &           & 11.19 (1.89)                          & \\		\hline 
			\multirow{13}{*}{$\tau=0.5$} & 	\multirow{11}{*}{Covariate}	 	       & Age (84\%)           &           & Age  (86\%)          &           &  Age (86\%)           &           & Age (84\%)                &  \\
			& & Continine (67\%)      &           & Continine (40\%)    &           & Continine (41\%)     &           & ILMN$\_$2125395 (33\%)            &  \\
			& & ILMN$\_$1759423 (26\%)        &           & ILMN$\_$1759423 (28\%)      &           & ILMN$\_$1759423 (24\%)       &           & ILMN$\_$1759423 (24\% )           &  \\
			& & ILMN$\_$1683806 (24\%)        &           &ILMN$\_$1683806 (19\%)      &           & ILMN$\_$1683806 (21\%)       &           & ILMN$\_$1683806 (21\%)            &  \\
			& & ILMN$\_$1779147 (24\%)        &           & ILMN$\_$1697433 (17\%)      &           & ILMN$\_$2051113 (16\%)       &           & {\it Continine} (12\%)    &  \\
			& & Parity (22\%)        &            &                     &           &                      &           &                           &  \\
			& & ILMN$\_$1738033(19\%)         &           &                     &           &                      &           &                           &  \\
			& & ILMN$\_$1739429  (19\%)        &          &                     &           &                      &           &                           &  \\
			& & ILMN$\_$1752218 (17\%)        &           &                     &           &                      &           &                           &  \\ 
			\cline{2-10}
			& QPE	& 0.223 (0.056)         &           &  0.214 (0.058)      &           & 0.217 (0.060)        &           & 0.228 (0.062)                          & \\
			& MS 		    & 12.17 (1.17)          &           &  12.26 (1.22)       &           & 12.07 (1.41)         &           & 11.89 (1.83)                          & \\		\hline
			
		\end{tabular}		
	}
\end{table}

\newpage 

\newpage

\section{Concluding Remarks}

In this paper, we have developed a flexible feature screening procedure for nonparametric additive quantile regression models in the ultra-high dimensional setting. We first use B-spline basis functions to approximate the nonparametric components. Then we propose an additive quantile forward screening method to subsequently screen important covariates. Finally, we apply a modified quantile Bayesian information criterion to the screened set for best subset selection. Compared to existing screening methods for parametric quantile models, our proposed procedure 
is more flexible and more robust to model misspecification. In addition, our procedure can  
successfully select important covariates when the variables are highly correlated, identify covariates that contribute to the quantiles conditionally but not marginally, and detect heterogeneous associations. 
These advantages of our proposed procedure are demonstrated through simulation studies and a real data example.

We have focused on the nonparametric additive quantile regression models in this paper. Recently, \cite{sherwood2016} proposed a penalized variable selection method for partially linear additive quantile regression models. Their approach applies penalization only on the linearly entered variables and assumes that the number of nonparametric components is finite. Our feature screening method can be easily extended to partially linear additive quantile regression models in ultra-high dimension where both the number of linearly entered variables and the number of nonparametric components can grow 
exponentially with the sample size. In addition, we take $q_n=[n^{1/5}]$ in our simulation studies. The choice of $q_n$ may also affect the numerical performance of our method in practice.  One can choose the optimal value of $q_n$ using cross-validation or other data-driven approaches. 
These possible extensions are beyond the scope of the current paper and will be an interesting topic for future research.

\section*{Acknowledgments}
The authors sincerely thank the Editor, Associate Editor, and anonymous referees for their valuable and constructive comments.

\section*{Disclosure statement}
No potential conflict of interest was reported by the authors.

\section*{Funding}
Li was supported by the 2022–2023 RSCA Award Program at California State University, Fullerton, United States.



\section*{Appendix A: Proofs of Theorems \ref{Th1}-\ref{Th4}}\label{AppA}

\renewcommand{\theequation}{A.\arabic{equation}}
\setcounter{equation}{0}
\setcounter{section}{0}
\renewcommand{\thesubsection}{A.\arabic{subsection}}

Recall that  $\hat{\btheta}_{\mathcal{S}}=\argmin_{{\btheta}_{\mathcal{S}}}n^{-1}\sum\limits_{i=1}^n\rho_{\tau}\left(Y_i-\bz_{i, \,\mathcal{S}}^T\btheta_{\mathcal{S}}\right)$. Define
$\btheta_{0{\mathcal{S}}}=\argmin_{{\btheta}_{\mathcal{S}}}E\left[\rho_{\tau}\left(Y-\bz_{\mathcal{S}}^T\btheta_{\mathcal{S}}\right)\right]$. Then $\hat{\btheta}_{\mathcal{S}}$ and 
$\btheta_{0\mathcal{S}}$ are the minimizers of
$g_n(\btheta_{\mathcal{S}})
=n^{-1}\sum\limits_{i=1}^n\left[\rho_{\tau}\left(Y_i-\bz_{i, \,\mathcal{S}}^T\btheta_{\mathcal{S}}\right)-\rho_{\tau}(Y_i)\right]
\,\,\mbox{and}\,\,
g(\btheta_{\mathcal{S}})
=E\left[\rho_{\tau}\left(Y-\bz_{\mathcal{S}}^T\btheta_{\mathcal{S}}\right)-\rho_{\tau}(Y)\right]$, respectively.  
Before proving our Theorems~\ref{Th1}-\ref{Th4}, we present three lemmas. Lemma \ref{Lemma: Hoeffding} is known as Hoeffding's lemma (see Lemma A in Section 5.6.1 of \cite{serfling1980approximation}). 
Lemmas~\ref{lem: gdiff-bound} and \ref{lem: theta-diff-bound}
are used in the proofs of Theorem \ref{Th1}.  The result in Theorem \ref{Th1} is the key to prove Theorem \ref{Th2}.
The proofs of lemmas~\ref{lem: gdiff-bound} and~\ref{lem: theta-diff-bound} are given in Appendix B. As mentioned in Remark 1 in Section~\ref{sec: theory}, when we let $q_n=O\left(n^{1/(2d+1)}\right)$, which is the optimal rate for B-spline approximation, and take $r_n=O(n^{\omega})$, the assumption  $r_n^{1/2}q_n^{-d}n^{\alpha}=o(1)$ will hold under the requirement $\omega<2d/(2d+1)-2\alpha$ for  $0<\alpha<d/(2d+1)$.

\begin{lemma}\label{Lemma: Hoeffding}
	Let $W$ be a real-valued random variable  with $E(W)=\mu_W$.  If $P(a\leq W\leq b)=1$ for some $a, b\in R$, 
	then $E\left\{\exp[t(W-\mu_W)]\right\}\leq \exp[t^2(b-a)^2/8]$ for any $t>0$.
\end{lemma}

\begin{lemma}\label{lem: gdiff-bound}
	Assume that $r_n^{1/2}q_n^{-d}n^{\alpha}=o(1)$. Then, for any constant $C>0$, there exists a positive constant $C_4$ such that
	\begin{align*}
	\inf_{\|\btheta_{\mathcal{S}}-\btheta_{0\mathcal{S}}\|=Cq_n^{-1/2}n^{-\alpha}}
	[g(\btheta_{\mathcal{S}})-g(\btheta_{0\mathcal{S}})]\geq 2C_4r_nn^{-2\alpha}
	\end{align*}
	for all $n$ sufficiently large.
\end{lemma}

\begin{lemma}\label{lem: theta-diff-bound}
	Assume that $r_n^{1/2}q_n^{-d}n^{\alpha}=o(1)$. Then, for any constant $C>0$, there exist some positive constants $C_5$ and $C_6$ such that 
	\begin{align*}
	P\left(\|\hat{\btheta}_{\mathcal{S}}-\btheta_{0\mathcal{S}}\|\geq Cq_n^{-1/2}n^{-\alpha}\right)
	\leq 2\exp\left(-C_5r_n^2n^{1-4\alpha}\right)+\exp\left(-C_6r_nn^{1-2\alpha}\right)
	\end{align*}
	for all $n$ sufficiently large.
\end{lemma}

\vspace{4mm}
\noindent{\bf Proof of Theorem \ref{Th1}}.
Recall that $\|\widehat{d}_k(\mathcal{S})\|
=n^{-1}\sum\limits_{i=1}^n\hat{d}_k^2(X_{ik}|\mathcal{S})$
and $\|d_k(\mathcal{S})\|={\rm E}[d^2_k(t|\mathcal{S})]$
with 
\begin{align*}
\hat{d}_k(t|\mathcal{S})
=n^{-1}\sum_{i=1}^n\left[\tau - I\left\{Y_i< \widehat{Q}_{\tau}(Y|\bx_{i,\,\mathcal{S}})\right\}\right]I(X_{ik}<t) 
\end{align*}
and $d_k(t|\mathcal{S})={\rm E}\Big(\left[\tau - I\left\{Y<Q_{\tau}(Y | \bx_{\mathcal{S}})\right\}\right]I(X_k<t)\Big)$
where $\widehat{Q}_{\tau}(Y|\bx_{i,\,\mathcal{S}})
=\bz_{i, \,\mathcal{S}}^T\hat{\btheta}_{\mathcal{S}}$. 
Let $e(\mathcal{S})=Y-Q_{\tau}(Y|\,\bx_{\mathcal{S}})$, 
$e_i(\mathcal{S})=Y_i-Q_{\tau}(Y_i|\,\bx_{i, \, \mathcal{S}})$ and 
$\hat{e}_i(\mathcal{S})=Y_i-\widehat{Q}_{\tau}(Y_i|\,\bx_{i, \, \mathcal{S}})$. Then we have
$\hat{d}_k(t|\mathcal{S})
=n^{-1}\sum\limits_{i=1}^n\left[\tau - I\{\hat{e}_i(\mathcal{S})< 0 \}\right]I(X_{ik}<t)
\,\,\mbox{and}
\,\,
d_k(t|\mathcal{S})
={\rm E}\Big(\left[\tau - I\{e_i(\mathcal{S})< 0 \}\right]I(X_k<t)\Big)$. 
In addition, we define
\begin{align*}
\tilde{d}_{k}(t|\mathcal{S})=n^{-1}\sum\limits_{i=1}^n\left[\tau-I\{e_i(\mathcal{S})<0\}\right]I(X_{ik}<t)
\,\,\mbox{and}
\,\,
\|\tilde{d}_{k}(\mathcal{S})\|= n^{-1}\sum\limits_{i=1}^n \tilde{d}_{k}^2(X_{ik}|\mathcal{S}).
\end{align*}
An application of the triangle inequality entails
\begin{align}\label{eq: B14}
\big|\|\widehat{d}_{k}(\mathcal{S})\|-\|d_{k}(\mathcal{S})\|\big| 
\leq
\big|\|\widehat{d}_{k}(\mathcal{S})\|-\|\tilde{d}_{k}(\mathcal{S})\|\big| 
+\big|\|\tilde{d}_{k}(\mathcal{S})\|-\|d_{k}(\mathcal{S})\|\big|.
\end{align}
In what follows, we will provide details on deriving an exponential tail probability bound for each term on the right hand side of \eqref{eq: B14}.
To enhance readability, we split the proof into two steps.

{\bf Step 1}.  We start with the first term $ \big|\|\widehat{d}_{k}(\mathcal{S})\|-\|\tilde{d}_{k}(\mathcal{S})\|\big|$. 
After some calculations, we have 
\begin{align}\label{eq: d-tilde-equiv}
\|\tilde{d}_{k}(\mathcal{S})\|
= \frac{(n-1)(n-2)}{n^2} \left[\frac{1}{n-2}\tilde{D}_{k1}(\mathcal{S})+\tilde{D}_{k2}(\mathcal{S})\right],
\end{align}
where
\begin{align}\label{eq: Dtilde-k1}
\tilde{D}_{k1}(\mathcal{S}) 
=& \frac{2}{n(n-1)}\sum_{i<j} \frac{1}{2}
\left\{[\tau-I\{e_i(\mathcal{S})<0\}]^2I(X_{ik}<X_{jk})
+[\tau-I\{e_j(\mathcal{S})<0\}]^2I(X_{jk}<X_{ik})
\right\} \nonumber\\
\overset{\Delta}{=} & \frac{2}{n(n-1)}\sum_{i<j} \varphi_1(X_{ik}, Y_i; X_{jk}, Y_j; \mathcal{S}),
\end{align}
and
\begin{align}\label{eq: Dtilde-k2}
 &\tilde{D}_{k2}(\mathcal{S})  \nonumber\\
=& \frac{6}{n(n-1)(n-2)}\sum_{i<j<\ell} \frac{1}{3}
\left\{[\tau-I\{e_i(\mathcal{S})<0\}] [\tau-I\{e_j(\mathcal{S})<0\}] I(X_{ik}<X_{\ell k})I(X_{jk}<X_{\ell k})\right. \nonumber\\
&    \quad\quad\quad\quad \quad\quad\quad\quad\quad\quad +[\tau-I\{e_j(\mathcal{S})<0\}] [\tau-I\{e_{\ell}(\mathcal{S})<0\}] 
I(X_{jk}<X_{ik})I(X_{\ell k}<X_{ik}) \nonumber\\
&    \quad\quad\quad\quad\quad\quad\quad\quad\quad\quad  \left.+[\tau-I\{e_{\ell}(\mathcal{S})<0\}] [\tau-I\{e_i(\mathcal{S})<0\}] 
I(X_{\ell k}<X_{jk})I(X_{ik}<X_{jk})\right\}  \nonumber\\
\overset{\Delta}{=} & \frac{6}{n(n-1)(n-2)}\sum_{i<j<\ell} \varphi_2(X_{ik}, Y_i; X_{jk}, Y_j; X_{\ell k}, Y_{\ell}; \mathcal{S}),
\end{align}
where the definitions of the kernels $\varphi_1$ and $\varphi_2$ are clear from the context above.
Obviously, both $\tilde{D}_{k1}(\mathcal{S}) $ and $\tilde{D}_{k2}(\mathcal{S})$ are $U$-statistics with the kernels $\varphi_1$ and $\varphi_2$, respectively.
Similarly, we can re-write $\|\widehat{d}_{k}(\mathcal{S})\|$ as 
\begin{align}\label{eq: dk-hat-equiv}
\|\widehat{d}_{k}(\mathcal{S})\|
= \frac{(n-1)(n-2)}{n^2} \left[\frac{1}{n-2}\widehat{D}_{k1}(\mathcal{S})+\widehat{D}_{k2}(\mathcal{S})\right],
\end{align}
where $\widehat{D}_{k1}(\mathcal{S}) $ and $\widehat{D}_{k2}(\mathcal{S}) $ are obtained by replacing $e_i(\mathcal{S}), e_j(\mathcal{S})$ and $e_{\ell}(\mathcal{S})$ in $\tilde{D}_{k1}$ and $\tilde{D}_{k2}$ with $\hat{e}_i(\mathcal{S}), \hat{e}_j(\mathcal{S})$ and $\hat{e}_{\ell}(\mathcal{S})$, respectively. 
Combining \eqref{eq: d-tilde-equiv} with \eqref{eq: dk-hat-equiv} leads to
\begin{align}\label{eq: dk-hat-dk-tilde}
&\big|\|\widehat{d}_{k}(\mathcal{S})\|-\|\tilde{d}_{k}(\mathcal{S})\|\big|
=
\left|\frac{n-1}{n^2} \left[\widehat{D}_{k1}(\mathcal{S})-\tilde{D}_{k1}(\mathcal{S})\right]
+ \frac{(n-1)(n-2)}{n^2} \left[\widehat{D}_{k1}(\mathcal{S})-\tilde{D}_{k2}(\mathcal{S})\right]\right| \nonumber\\
\leq& n^{-1}\left|\widehat{D}_{k1}(\mathcal{S})-\tilde{D}_{k1}(\mathcal{S})\right|
+ \left|\widehat{D}_{k2}(\mathcal{S})-\tilde{D}_{k2}(\mathcal{S})\right|.
\end{align}
It follows from the defintions of $\widehat{D}_{k1}(\mathcal{S})$ and
$\tilde{D}_{k1}(\mathcal{S})$, the fact $0<\tau<1$, and the boundedness of the indicator functions that  
$0\leq \widehat{D}_{k1}(\mathcal{S})\leq 1$
and $0\leq \tilde{D}_{k1}(\mathcal{S})\leq 1$.
Thus, for any $\delta>0$, $n^{-1}\left|\widehat{D}_{k1}(\mathcal{S})-\tilde{D}_{k1}(\mathcal{S})\right|<\delta$ holds for all $1\leq k\leq p$ and all $n$ sufficiently large. This together with \eqref{eq: dk-hat-dk-tilde} yields that
\begin{align}\label{eq: dk-hat-bound1}
P\left(\big|\|\widehat{d}_{k}(\mathcal{S})\|-\|\tilde{d}_{k}(\mathcal{S})\|\big|\geq 2\delta\right)
\leq P\left(\left|\widehat{D}_{k2}(\mathcal{S})-\tilde{D}_{k2}(\mathcal{S})\right|\geq \delta\right)
\end{align}
for any $\delta>0$ and all $n$ sufficiently large. 
Thus, it suffices to bound the probability
$P\left(\left|\widehat{D}_{k2}(\mathcal{S})-\tilde{D}_{k2}(\mathcal{S})\right|\geq \delta\right)$.

It follows from the triangle inequality and the boundedness of the indicator functions that 
we have
\begin{align}\label{eq: Dhat-k2-Dtilde-k2-bound1}
|\widehat{D}_{k2}(\mathcal{S}) -\tilde{D}_{k2}(\mathcal{S})|
\leq 2(1+\tau)n^{-1}\sum_{i=1}^n\left| I\{\hat{e}_i(\mathcal{S})<0\}-I\{e_i(\mathcal{S})<0\}\right|.
\end{align}
Hereafter our analysis will be conditional on the event $\Omega_1=\left\{\|\hat{\btheta}_{\mathcal{S}}-\btheta_{0\mathcal{S}}\|\leq c_6^{-1}Cq_n^{-1/2}n^{-\alpha}\right\}$, where
$c_6=1+5(1+\tau)f_{\max}$ and $f_{\max}=\sup\left\{ f_{Y|\bx_{\mathcal{S}}}(Q_{\tau}(Y|\,\bx_{\mathcal{S}}))\right\}$.
Conditional on the event $\Omega_1$, using the fact
of $\|B_{j\ell}(\cdot)\|_{\infty}\leq 1$ and the Cauchy-Schwarz inequality, we have 
$\left|\bz_{i, \,\mathcal{S}}^T(\hat{\btheta}_{\mathcal{S}}-\btheta_{0\mathcal{S}})\right|
\leq c_6^{-1}Cr_n^{1/2}n^{-\alpha}$. 
Following similar arguments in the proof of Lemma 3.1 in \cite{he2013}, there exist a positive constant $c_5$ such that $\left|Q_{\tau}(Y|\bx_{\mathcal{S}})-\bz_{i, \,\mathcal{S}}^T\btheta_{0\mathcal{S}}\right|
\leq c_5r_nq_n^{-d}$.
An application of the triangle inequality entails
\begin{align*}
&\left|\widehat{Q}_{\tau}(Y_i|\,\bx_{i, \, \mathcal{S}})-Q_{\tau}(Y_i|\,\bx_{i, \, \mathcal{S}})\right|
=\left|\bz_{i, \,\mathcal{S}}^T(\hat{\btheta}_{\mathcal{S}}-\btheta_{0\mathcal{S}})-\left[Q_{\tau}(Y_i|\,\bx_{i, \, \mathcal{S}})-\bz_{i, \,\mathcal{S}}^T\btheta_{0\mathcal{S}}\right]\right|\nonumber\\
\leq & \left|\bz_{i, \,\mathcal{S}}^T(\hat{\btheta}_{\mathcal{S}}-\btheta_{0\mathcal{S}})\right|
+\left|Q_{\tau}(Y_i|\,\bx_{i, \, \mathcal{S}})-\bz_{i, \,\mathcal{S}}^T\btheta_{0\mathcal{S}}\right|
\leq c_6^{-1}Cr_n^{1/2}n^{-\alpha} + c_5r_nq_n^{-d}
\leq 2c_6^{-1}Cr_n^{1/2}n^{-\alpha},
\end{align*}
for sufficiently large $n$, 
where the last equality follows from
the fact $r_n^{1/2}q_n^{-d}n^{\alpha}=o(1)$ by noting that  $q_n=O\left(n^{1/(2d+1)}\right)$,  $r_n=O(n^{\omega})$, and $\omega<2d/(2d+1)-2\alpha$ for  $0<\alpha<d/(2d+1)$.

Using the fact that $\sup_{s: |s-t|\leq \epsilon}|I\{Y<s\}-I\{Y<t\}|\leq I\{t-\epsilon\leq Y\leq t+\epsilon\}$
for any $\epsilon>0$, we have
\begin{align*}
&\big| I\{\hat{e}_i(\mathcal{S})<0\}-I\{e_i(\mathcal{S})<0\}\big|
=\big| I\left\{Y_i<\widehat{Q}_{\tau}(Y_i|\,\bx_{i, \, \mathcal{S}})\right\}-I\left\{Y_i<Q_{\tau}(Y_i|\,\bx_{i, \, \mathcal{S}})\right\}\big| \nonumber\\
\leq & I\left\{Q_{\tau}(Y_i|\,\bx_{i, \, \mathcal{S}})-2c_6^{-1}Cr_n^{1/2}n^{-\alpha}\leq Y_i\leq Q_{\tau}(Y_i|\,\bx_{i, \, \mathcal{S}})+2c_6^{-1}Cr_n^{1/2}n^{-\alpha}\right\}\nonumber\\
=& I\left\{-2c_6^{-1}Cr_n^{1/2}n^{-\alpha}\leq e_i(\mathcal{S})\leq 2c_6^{-1}Cr_n^{1/2}n^{-\alpha}\right\}.
\end{align*}
Combining this with \eqref{eq: Dhat-k2-Dtilde-k2-bound1} leads to
\begin{align}\label{eq: Dhat-k2-Dtilde-k2}
|\widehat{D}_{k2}(\mathcal{S}) -\tilde{D}_{k2}(\mathcal{S})|
\leq 2(1+\tau) n^{-1}\sum_{i=1}^n
I\left\{-2c_6^{-1}Cr_n^{1/2}n^{-\alpha}\leq e_i(\mathcal{S})\leq 2c_6^{-1}Cr_n^{1/2}n^{-\alpha}\right\}.
\end{align}
Let 
$\mu=E\left(I\left\{-2c_6^{-1}Cr_n^{1/2}n^{-\alpha}\leq e(\mathcal{S})\leq 2c_6^{-1}Cr_n^{1/2}n^{-\alpha}\right\}\right)$
with
$e(\mathcal{S})=Y-Q_{\tau}(Y|\,\bx_{\mathcal{S}})$. By the Taylor expansions, we further have 
\begin{align*}
\mu=2f_{Y|\bx_{\mathcal{S}}}(Q_{\tau}(Y|\,\bx_{\mathcal{S}}))(2c_6^{-1}Cr_n^{1/2}n^{-\alpha})+\left[f'_{Y|\bx_{\mathcal{S}}}(\zeta_1)-f'_{Y|\bx_{\mathcal{S}}}(\zeta_2)\right](2c_6^{-1}Cr_n^{1/2}n^{-\alpha})^2/2,
\end{align*}
where $\zeta_1$ and $\zeta_2$ are in the ($2c_6^{-1}Cr_n^{1/2}n^{-\alpha}$)-neighborhood of $Q_{\tau}(Y|\,\bx_{\mathcal{S}})$. Under Condition \ref{con: fy}, there exists some positive constant $c_7$ such that $\left|f'_{Y|\bx_{\mathcal{S}}}(\zeta_1)-f'_{Y|\bx_{\mathcal{S}}}(\zeta_2)\right|\leq c_7$. This leads to 
\begin{align*}
\mu\leq 4c_6^{-1}Cf_{\max}r_n^{1/2}n^{-\alpha}+c_6^{-1}Cr_n^{1/2}n^{-\alpha}/(1+\tau),
\end{align*}
by noting that 
$2c_7(1+\tau)(Cr_n^{1/2}n^{-\alpha})\leq c_6$ for sufficiently large $n$.
Recall that $c_6=1+5(1+\tau)f_{\max}$. 
We further have
\begin{align*}
\frac{Cr_n^{1/2}n^{-\alpha}}{1+\tau}-\mu
\geq & \frac{c_6^{-1}Cr_n^{1/2}n^{-\alpha}}{1+\tau}\left[1+5(1+\tau)f_{\max}\right]-4c_6^{-1}Cf_{\max}r_n^{1/2}n^{-\alpha}-\frac{c_6^{-1}Cr_n^{1/2}n^{-\alpha}}{1+\tau} \nonumber\\
=& c_6^{-1}Cf_{\max}r_n^{1/2}n^{-\alpha}.
\end{align*}
This together with \eqref{eq: Dhat-k2-Dtilde-k2} yields 
\begin{align*}
& P\left(|\widehat{D}_{k2}(\mathcal{S}) -\tilde{D}_{k2}(\mathcal{S})|\geq Cr_n^{1/2}n^{-\alpha}|\Omega_1\right) \nonumber\\
\leq & P\left(n^{-1}\sum_{i=1}^n
I\left\{-2c_6^{-1}Cr_n^{1/2}n^{-\alpha}\leq e_i(\mathcal{S})\leq 2c_6^{-1}Cr_n^{1/2}n^{-\alpha}\right\}\geq \frac{Cr_n^{1/2}n^{-\alpha}}{1+\tau}\right)\nonumber\\
\leq & P\left(n^{-1}\sum_{i=1}^n
I\left\{-2c_6^{-1}Cr_n^{1/2}n^{-\alpha}\leq e_i(\mathcal{S})\leq 2c_6^{-1}Cr_n^{1/2}n^{-\alpha}\right\}-\mu\geq c_6^{-1}Cf_{\max}r_n^{1/2}n^{-\alpha}\right)\nonumber\\
\leq & \exp\left(-2c_6^{-2}C^2f^2_{\max}r_nn^{1-2\alpha}\right),
\end{align*}
where the last inequality follows from Hoeffding's inequality \citep{hoeffding1963probability}.  This together with Lemma \ref{lem: theta-diff-bound} and \eqref{eq: dk-hat-bound1} entails
\begin{align}\label{eq: term1-bound}
& P\left(\big|\|\widehat{d}_{k}(\mathcal{S})\|-\|\tilde{d}_{k}(\mathcal{S})\|\big|\geq 2Cr_n^{1/2}n^{-\alpha}\right) 
\leq P\left(|\widehat{D}_{k2}(\mathcal{S}) -\tilde{D}_{k2}(\mathcal{S})|\geq Cr_n^{1/2}n^{-\alpha}|\Omega_1\right)+P(\Omega_1^c)\nonumber\\
\leq &
\exp\left(-2c_6^{-2}C^2f^2_{\max}r_nn^{1-2\alpha}\right)+ 2\exp\left(-C_5r_n^2n^{1-4\alpha}\right)+\exp\left(-C_6r_nn^{1-2\alpha}\right)\nonumber\\
\leq &  2\exp\left(-C_5r_n^2n^{1-4\alpha}\right)+2\exp\left(-C_7r_nn^{1-2\alpha}\right)
\end{align}
for some positive constant $C_7$.

{\bf Step 2} We will establish the probability bound for $\big|\|\tilde{d}_{k}(\mathcal{S})\|-\|d_{k}(\mathcal{S})\|\big|$. Recall that $\|d_k(\mathcal{S})\|={\rm E}[d^2_k(t|\mathcal{S})]$
with $d_k(t|\mathcal{S})={\rm E}\Big(\left[\tau - I\left\{e(\mathcal{S})<0\right\}\right]I(X_k<t)\Big)
$. With the definition of the kernel $\varphi_2$ in \eqref{eq: Dtilde-k2}, we can show that $\|d_k(\mathcal{S})\|=E\left[\varphi_2(X_{ik}, Y_i; X_{jk}, Y_j; X_{\ell k}, Y_{\ell}; \mathcal{S})\right]=E\left[\tilde{D}_{k2}(\mathcal{S})\right]$. This together with \eqref{eq: d-tilde-equiv}
and the triangle inequality entails
\begin{align}\label{eq: term2-bound1}
&\left|\|\tilde{d}_{k}(\mathcal{S})\|-\|d_{k}(\mathcal{S})\|\right| \nonumber\\ 
=& \left|\frac{n-1}{n^2}\tilde{D}_{k1}(\mathcal{S})
+\frac{(n-1)(n-2)}{n^2}\left[\tilde{D}_{k2}(\mathcal{S})-E(\tilde{D}_{k2}(\mathcal{S})) \right]
-\frac{3n-2}{n^2}E(\tilde{D}_{k2}(\mathcal{S}))\right|\nonumber\\
\leq & n^{-1}|\tilde{D}_{k1}(\mathcal{S})|+\left|\tilde{D}_{k2}(\mathcal{S})-E(\tilde{D}_{k2}(\mathcal{S})) \right|
+3n^{-1}|E(\tilde{D}_{k2}(\mathcal{S}))|.
\end{align}
It follows from the defintions of $\tilde{D}_{k1}(\mathcal{S})$ in \eqref{eq: Dtilde-k1} and
$\tilde{D}_{k2}(\mathcal{S})$ in \eqref{eq: Dtilde-k2}, the fact $0<\tau<1$, and the boundedness of the indicator functions that  
$0\leq \tilde{D}_{k1}(\mathcal{S})\leq 1$
and $-1\leq \tilde{D}_{k2}(\mathcal{S})\leq 1$. 
Thus, 
for any $\delta>0$, we have $0\leq n^{-1}|\tilde{D}_{k1}(\mathcal{S})|<\delta$ and $0\leq 3n^{-1}|E(\tilde{D}_{k2}(\mathcal{S}))|< \delta$ hold uniformly for all $1\leq k\leq p$ and all $n$ sufficiently large.
Combining this with \eqref{eq: term2-bound1} yields 
\begin{align}\label{eq: term2-bound2}
P\left( \left|\|\tilde{d}_{k}(\mathcal{S})\|-\|d_{k}(\mathcal{S})\|\right|\geq 3\delta\right)
\leq P\left(\left|\tilde{D}_{k2}(\mathcal{S})-E(\tilde{D}_{k2}(\mathcal{S})) \right|\geq \delta\right)
\end{align}
for any $\delta>0$. Thus it is sufficient to
the probability on the right hand side above.

For any $\delta>0$, by the Markov's inequality, we have
\begin{align}\label{eq: tilde-Dm2}
P\left(\tilde{D}_{k2}(\mathcal{S})-E(\tilde{D}_{k2}(\mathcal{S})) \geq \delta\right)
\leq \exp(-t\delta)\exp\left[-t E(\tilde{D}_{k2}(\mathcal{S})) \right]E\left\{\exp(t\tilde{D}_{k2}(\mathcal{S}))\right\}
\end{align}
for any $t>0$.   
Recall that $\varphi_2$ is the kernel of the $U$-statistic $\tilde{D}_{k2}(\mathcal{S})$.  
According to the theory of $U$-statistics \cite[sec. 5.1.6]{serfling1980approximation}, any $U$-statistic can be represented as an average of averages of independent and identically distributed random variables.  
This representation gives
\begin{align*}
\tilde{D}_{k2}(\mathcal{S})=(n!)^{-1}\sum_{n!}D(X_{i_1,\,m}, Y_{i_1}; \cdots, X_{i_n, \,m}, Y_{i_n}; \mathcal{S})
\end{align*}
where $\sum_{n!}$ denotes the summation over all possible permutations $(i_1,\cdots, i_n)$ of $(1, \cdots, n)$, and 
\begin{align}\label{eq:D-def}
&D(X_{1k}, Y_1; \cdots; X_{nk}, Y_n; \mathcal{S})\nonumber \\
=& n_1^{-1}[\varphi_2(X_{1k}, Y_1; X_{2k}, Y_2; X_{3k}, Y_{3}; \mathcal{S}) 
+\varphi_2(X_{4k}, Y_4; X_{5k}, Y_5; X_{6k}, Y_{6}; \mathcal{S}) \nonumber \\
& \quad\quad     +\cdots+ \varphi_2(X_{3n_1-2, \,k}, Y_{3n_1-2}; X_{3n_1-1, \,k}, Y_{3n_1-1}; X_{3n_1, \,k}, Y_{3n_1}; \mathcal{S}) ]
\end{align}
with $n_1=\lfloor n/3 \rfloor$ the integer part of $n/3$.
An application of Jensen's inequality yields that 
\begin{align*}
& E\left\{\exp[t\tilde{D}_{k2}(\mathcal{S})]\right\}
= E\left\{\exp\left[(n!)^{-1}\sum _{n!}tD(X_{i_1,\,k}, Y_{i_1}; \cdots; X_{i_n, \,k}, Y_{i_n}; \mathcal{S})\right]\right\} \\
\leq & E\left\{(n!)^{-1}\sum _{n!}\exp\left[tD(X_{i_1,\,k}, Y_{i_1}; \cdots; X_{i_n, \,k}, Y_{i_n}; \mathcal{S})\right]\right\}\\ 
=&  (n!)^{-1}\sum _{n!}E\left\{\exp\left[tD(X_{i_1,\,k}, Y_{i_1}; \cdots; X_{i_n, \,k}, Y_{i_n}; \mathcal{S})\right]\right\} \\
=& E\left\{\exp\left[tD(X_{1k}, Y_{1}; \cdots; X_{nk}, Y_{n}; \mathcal{S})\right]\right\} 
= E^{n_1}\left\{\exp\left[tn_1^{-1}\varphi_2(X_{1k}, Y_1; X_{2k}, Y_2; X_{3k}, Y_{3}; \mathcal{S}) \right]\right\}
\end{align*}
for any $t>0$, where the last equality follows from \eqref{eq:D-def}.  The above inequality together with (\ref{eq: tilde-Dm2}) leads to 
\begin{align*}
P\left(\tilde{D}_{k2}(\mathcal{S})-E(\tilde{D}_{k2}(\mathcal{S})) \geq \delta\right)
\leq \exp(-t\delta)E^{n_1}\left\{\exp\left[tn_1^{-1}\tilde{\varphi}_2(X_{1k}, Y_1; X_{2k}, Y_2; X_{3k}, Y_{3}; \mathcal{S}) \right]\right\}
\end{align*}
with $\tilde{\varphi}_2(X_{1k}, Y_1; X_{2k}, Y_2; X_{3k}, Y_{3}; \mathcal{S})=\varphi_2(X_{1k}, Y_1; X_{2k}, Y_2; X_{3k}, Y_{3}; \mathcal{S})-E(\tilde{D}_{k2}(\mathcal{S}))$.

Note that 
$-1\leq \varphi_2(X_{1m}, Y_1; X_{2m}, Y_2; X_{3m}, Y_{3}; \mathcal{S})\leq 1$ and $E\left[\varphi_2(X_{1k}, Y_1; X_{2k}, Y_2; X_{3k}, Y_{3}; \mathcal{S})\right]=E(\tilde{D}_{k2}(\mathcal{S}))$.
It follows from Lemma \ref{Lemma: Hoeffding} that 
$E\left\{\exp\left[tn_1^{-1}\tilde{\varphi}_2(X_{1k}, Y_1; X_{2k}, Y_2; X_{3k}, Y_{3}; \mathcal{S}) \right]\right\}\leq \exp(2^{-1}n_1^{-1}t^2)$
and then $	P\left(\tilde{D}_{k2}(\mathcal{S})-E(\tilde{D}_{k2}(\mathcal{S})) \geq \delta\right)
\leq \exp\left(-t\delta+2^{-1}n_1^{-1}t^2\right)$
for any $t>0$ and $\delta>0$.
Minimizing the right-hand side above with respect to $t$ gives
\begin{align*}
P\left(\tilde{D}_{k2}(\mathcal{S})-E(\tilde{D}_{k2}(\mathcal{S})) \geq\delta\right)
\leq \exp\left(-n_1\delta^2/2\right)
\end{align*}
for any $\delta>0$. Similarly, we can show that
$P\left(\tilde{D}_{k2}(\mathcal{S})-E(\tilde{D}_{k2}(\mathcal{S}))\leq-\delta\right)
\leq \exp\left(-n_1\delta^2/2\right)$ 
for any $\delta>0$. Therefore, it holds that
\begin{eqnarray*}
	P\left(\left|\tilde{D}_{k2}(\mathcal{S})-E(\tilde{D}_{k2}(\mathcal{S}))  \right| \geq \delta\right)
	\leq 2\exp\left(-n_1\delta^2/2\right).
\end{eqnarray*}
Recall that $n_1=\lfloor n/3 \rfloor$.  Then for 
$\delta=Cr_n^{1/2}n^{-\alpha}$ 
with any positive constant $C$, there exists some positive constant $C_8$ such that when $n$ is sufficiently large
\begin{eqnarray*}
	P\left(\left|\tilde{D}_{k2}(\mathcal{S})-E(\tilde{D}_{k2}(\mathcal{S})) \right| \geq Cr_n^{1/2}n^{-\alpha}\right)
	\leq 2\exp\left(-C_8r_nn^{1-2\alpha}\right)
\end{eqnarray*}
for all $1\leq k\leq p$.  This together with \eqref{eq: term2-bound2} yields
\begin{align}\label{eq: term2-bound}
P\left( \left|\|\tilde{d}_{k}(\mathcal{S})\|-\|d_{k}(\mathcal{S})\|\right|\geq 3Cr_n^{1/2}n^{-\alpha}\right)
\leq 2\exp\left(-C_8r_nn^{1-2\alpha}\right).
\end{align}

In view of \eqref{eq: B14}, \eqref{eq: term1-bound} and \eqref{eq: term2-bound}, we have
\begin{align}\label{eq-A16}
& P\left(\big|\|\widehat{d}_{k}(\mathcal{S})\|-\|d_{k}(\mathcal{S})\|\big|\geq 5Cr_n^{1/2}n^{-\alpha}\right)\nonumber\\
\leq & 2\exp\left(-C_5r_n^2n^{1-4\alpha}\right)+2\exp\left(-C_7r_nn^{1-2\alpha}\right)
+2\exp\left(-C_8r_nn^{1-2\alpha}\right)\nonumber\\
\leq & 2\exp\left(-c_2r_n^2n^{1-4\alpha}\right)+4\exp\left(-c_3r_nn^{1-2\alpha}\right)
\end{align}
for some positive constants $c_2$ and $c_3$. Thus, 
\begin{align*}
P\left(\max_{1\leq k\leq p}\big|\|\widehat{d}_{k}(\mathcal{S})\|-\|d_{k}(\mathcal{S})\|\big|\geq 5Cr_n^{1/2}n^{-\alpha}\right)
\leq  2p\exp\left(-c_2r_n^2n^{1-4\alpha}\right)+4p\exp\left(-c_3r_nn^{1-2\alpha}\right).
\end{align*}  
This concludes the proof of Theorem~\ref{Th1}.

\vspace{4mm}

\vspace{4mm}
\noindent{\bf Proof of Theorem \ref{Th2}}. On the event 
$	\Omega_2 
=\left\{\max_{k\in\mathcal{S}^*}\big|\|\widehat{d}_{k}(\mathcal{S})\|-\|d_{k}(\mathcal{S})\|\big|< C_0r_n^{1/2}n^{-\alpha}\right\}$,
it follows from Condition~\ref{con: min-signal} that 
$\|\widehat{d}_{k}(\mathcal{S})\|\geq \|d_{k}(\mathcal{S})\|-C_0r_n^{1/2}n^{-\alpha}\geq C_0r_n^{1/2}n^{-\alpha}$ for each $k\in \mathcal{S}^*$. Therefore, by the choice of $v_n=C_0r_n^{1/2}n^{-\alpha}$, we have $\mathcal{S}^{*} \subset  \widehat{\mathcal{S}}_{v_n}$ on the event $\Omega_2$. This indicates $P(\mathcal{S}^{*} \subset \widehat{\mathcal{S}}_{v_n}) \geq P(\Omega_2)$. Taking $C=C_0/5$ in~\eqref{eq-A16} and using the union bound and~\eqref{eq-A16}, we have
\begin{align*}
& P(\Omega_2)=1-P(\Omega_2^c)=1- P\left(\max_{k\in\mathcal{S}^*}\big|\|\widehat{d}_{k}(\mathcal{S})\|-\|d_{k}(\mathcal{S})\|\big|\geq C_0r_n^{1/2}n^{-\alpha}\right) \nonumber\\
\geq & 1-|\mathcal{S}^{*}|\left[2\exp\left(-c_2r_n^2n^{1-4\alpha}\right)+2\exp\left(-c_3r_nn^{1-2\alpha}\right)\right],
\end{align*}
This completes the proof of Theorem \ref{Th2}.

\vspace{4mm}

\vspace{4mm}
\noindent{\bf Proof of Theorem~\ref{Th: size}}. Under the assumption $\sum\limits_{k=1}^p\|d_{k}(\mathcal{S})\|=O(n^{\eta})$ for some $\eta>0$, 
the cardinality of the set $\left\{k: \|d_{k}(\mathcal{S})\|\geq 2^{-1}C_0r_n^{1/2}n^{-\alpha}\right\}$ cannot exceed $O(r_n^{-1/2}n^{\eta+\alpha})=O(n^{\eta+\alpha-\omega/2})$, where $r_n=O(n^{\omega})$ with 
$0\leq \omega< \min\{2\alpha,\, 2d/(2d+1)-2\alpha\}$. Furthermore, on the event 
$	\Omega_3=\left\{\max_{1\leq k\leq p}\big|\|\widehat{d}_{k}(\mathcal{S})\|-\|d_{k}(\mathcal{S})\|\big|< 2^{-1}C_0r_n^{1/2}n^{-\alpha}\right\}$,
we have
\begin{align*}
\big|\widehat{\mathcal{S}}_{v_n}\big|
=\big|\left\{k: \|\widehat{d}_{k}(\mathcal{S}^{(\ell -1)})\| 
\geq C_0r_n^{1/2}n^{-\alpha}\,\,\mbox{for}\,\,1\leq \ell\leq K_n \right\}\big|\\
\leq 	
\big|\left\{k: \|d_{k}(\mathcal{S})\|\geq 2^{-1}C_0r_n^{1/2}n^{-\alpha}\right\}\big|
\leq O(n^{\eta+\alpha-\omega/2}).
\end{align*}
This result, together with~\eqref{eq-A16} and the union bound, yields
\begin{align*}
&P\left(\big|\widehat{\mathcal{S}}_{v_n}\big|
\leq O(n^{\eta+\alpha-\omega/2})\right)
\geq P(\Omega_3) \\
=&
1-P(\Omega_3^c)=1- P\left(\max_{1\leq k\leq p}\big|\|\widehat{d}_{k}(\mathcal{S})\|-\|d_{k}(\mathcal{S})\|\big|\geq 2^{-1}C_0r_n^{1/2}n^{-\alpha}\right) \nonumber\\
&\geq  1-p\left[2\exp\left(-c_2r_n^2n^{1-4\alpha}\right)+2\exp\left(-c_3r_nn^{1-2\alpha}\right)\right].
\end{align*}
This completes the proof of Theorem~\ref{Th: size}.

\vspace{4mm}
\vspace{4mm}
\noindent{\bf Proof of Theorem~\ref{Th4}}. 
Recall that $\widehat{\mathcal{S}}_{\mathrm{QBIC}}=\mathcal{S}^{(\hat{\ell})}$. Let $\ell_{\min}=\min\limits_{1\leq \ell\leq K_n}\{\ell: \mathcal{S}^{*}\subset \mathcal{S}^{(\ell)}\}$.
It follows from Theorem~\ref{Th2} that $\ell_{\min}$ is well defined. 
Then, in order to prove Theorem~\ref{Th4}, it is sufficient to show that $P(\hat{\ell}\geq \ell_{\min})\to 1$ as $n\to\infty$, where $\hat{\ell}$ is given in \eqref{eq: ell_hat_def}. 
By Theorem~\ref{Th: size}, we have $K_n\leq O(n^{\eta+\alpha-\omega/2})$. This result, together with the definition of $\ell_{\min}$, yields that $\ell_{\min}\leq K_n\leq O(n^{\eta+\alpha-\omega/2})$. Write $R_i=\bz_i^{\top}\btheta^{*}-\mu_{\tau}-\sum_{j=1}^pg_{j,\,\tau}(X_{ij})$ for $i=1, \cdots, n$.
By (A.17) in~\cite{LeeNohPark2014}, we can choose a sequence of constants $\{L_n\}$ satisfying $L_n\to\infty$ and $L_n/C_n\to 0$ such that, with probability tending to one, 
\begin{align*}
&n^{-1}\left|\sum_{i=1}^n\rho_{\tau}\left(Y_i-\bz^{\top}_{i,\,\mathcal{S}^{(\ell_{\min})}}\hat{\btheta}_{\mathcal{S}^{(\ell_{\min})}}\right)-\rho_{\tau}(U_i-R_i)\right|\nonumber\\
=& 	n^{-1}\left|\sum_{i=1}^n\rho_{\tau}\left(U_i-\bz^{\top}_{i,\,\mathcal{S}^{(\ell_{\min})}}\left[\hat{\btheta}_{\mathcal{S}^{(\ell_{\min})}}-\btheta^{*}_{\mathcal{S}^{(\ell_{\min})}}\right]-R_i\right)-\rho_{\tau}(U_i-R_i)\right|\nonumber\\
\leq & n^{-1}L_nN_{\mathcal{S}^{(\ell_{\min})}}\log K_n
\leq \widetilde{C}_1n^{-1}L_nn^{\eta+\alpha-\omega/2}q_n\log n
\end{align*}
for some positive constant $\widetilde{C}_1$, where the last inequality follows from the facts that  
$N_{\mathcal{S}^{(\ell_{\min})}} =1+q_n\ell_{\min}$ and $\ell_{\min}\leq K_n\leq O(n^{\eta+\alpha-\omega/2})$.

By the fact $L_n/C_n\to 0$ and the assumption that $n^{\eta+\alpha-\omega/2}n^{-1}C_nq_n\log(n)=o(1)$, we have
\begin{align}\label{eq: A17}
n^{-1}\left|\sum_{i=1}^n\rho_{\tau}\left(Y_i-\bz^{\top}_{i,\,\mathcal{S}^{(\ell_{\min})}}\hat{\btheta}_{\mathcal{S}^{(\ell_{\min})}}\right)-\rho_{\tau}(U_i-R_i)\right|=o(1).
\end{align}
Since $\rho_{\tau}(u)=u[\tau-I(u<0)]=2^{-1}[|u|+(2\tau-1)u]$,
using the reverse triangle inequality yields 
\begin{align}\label{eq: rho-Lipschitz}
|\rho_{\tau}(u_1)-\rho_{\tau}(u_2)|
=& 2^{-1}\big||u_1|-|u_2|+(2\tau-1)(u_1-u_2)\big|%
\nonumber\\
\leq & 2^{-1}(1+|2\tau-1|)|u_1-u_2|
\leq |u_1-u_2| 
\end{align}
for any $u_1, u_2\in\mathbb{R}$. Thus, we have 
\begin{align}\label{eq: A19}
n^{-1}\left|\sum_{i=1}^n\left[\rho_{\tau}(U_i-R_i)-\rho_{\tau}(U_i)\right]\right|
\leq n^{-1}\sum_{i=1}^n\left|\rho_{\tau}(U_i-R_i)-\rho_{\tau}(U_i)\right|
\leq n^{-1}\sum_{i=1}^n |R_i|\leq \Delta^{*}=o(1).
\end{align}
Note that $\rho_{\tau}(u)=u[\tau-I(u<0)]\leq (\min\{1-\tau, \tau\})|u|\leq |u|$. Therefore, by the assumption that $E|U|<\infty$, we can show that $\widetilde{C}_2\leq E\rho_{\tau}(U)\leq \widetilde{C}_3$ for some constants $0<\widetilde{C}_2, \widetilde{C}_3<\infty$ and  $n^{-1}\sum_{i=1}^n\rho_{\tau}(U_i)$ convergences to $E\rho_{\tau}(U)$ in probability as $n\to\infty$. Combining these results with 
\eqref{eq: A17} and \eqref{eq: A19} entails that, with probability tending to one, 
\begin{align*}
\widetilde{C}_2+o(1)\leq n^{-1}\sum_{i=1}^n\rho_{\tau}\left(Y_i-\bz^{\top}_{i,\,\mathcal{S}^{(\ell_{\min})}}\hat{\btheta}_{\mathcal{S}^{(\ell_{\min})}}\right)\leq \widetilde{C}_3+o(1).
\end{align*}

Note that $\mathcal{S}^{(\ell)}$ for all $1\leq \ell<\ell_{\min}$ are underfitted models such that $\mathcal{S}^{*}\not\subset\mathcal{S}^{(\ell)}$ and $\mathcal{S}^{(\ell)}$ are nested. Using the same techniques as those used for the proof of (A.19) in~\cite{LeeNohPark2014},
we can show that there exists some positive constant $\widetilde{C}_4$ such that 
\begin{align}\label{eq: A21}
n^{-1}\sum_{i=1}^n\rho_{\tau}\left(Y_i-\bz^{\top}_{i,\,\mathcal{S}^{(\ell)}}\hat{\btheta}_{\mathcal{S}^{(\ell)}}\right)-n^{-1}\sum_{i=1}^n\rho_{\tau}\left(Y_i-\bz^{\top}_{i,\,\mathcal{S}^{(\ell_{\min})}}\hat{\btheta}_{\mathcal{S}^{(\ell_{\min})}}\right)\geq 2\widetilde{C}_4
\end{align}
with probability tending to one for any $1\leq \ell<\ell_{\min}$.
Therefore, we have
\begin{align*}
&\min_{1\leq \ell< \ell_{\min}}\mbox{QBIC}(\mathcal{S}^{(\ell)})-\mbox{QBIC}(\mathcal{S}^{(\ell_{\min})}) \\
=& \min_{1\leq \ell< \ell_{\min}}\left\{\log\left(1+\frac{n^{-1}\sum_{i=1}^n\rho_{\tau}\left(Y_i-\bz^{\top}_{i,\,\mathcal{S}^{(\ell)}}\hat{\btheta}_{\mathcal{S}^{(\ell)}}\right)-n^{-1}\sum_{i=1}^n\rho_{\tau}\left(Y_i-\bz^{\top}_{i,\,\mathcal{S}^{(\ell_{\min})}}\hat{\btheta}_{\mathcal{S}^{(\ell_{\min})}}\right)}{n^{-1}\sum_{i=1}^n\rho_{\tau}\left(Y_i-\bz^{\top}_{i,\,\mathcal{S}^{(\ell_{\min})}}\hat{\btheta}_{\mathcal{S}^{(\ell_{\min})}}\right)}\right)\right.\\
&\quad\quad\quad\quad\quad\left.+(N_{\mathcal{S}^{(\ell)}}-N_{\mathcal{S}^{(\ell_{\min})}})\frac{\log n}{2n}C_n\right\}\\
\geq & \min\left\{\log2, \, \frac{\widetilde{C}_4}{n^{-1}\sum_{i=1}^n\rho_{\tau}\left(Y_i-\bz^{\top}_{i,\,\mathcal{S}^{(\ell_{\min})}}\hat{\btheta}_{\mathcal{S}^{(\ell_{\min})}}\right)}\right\}-\widetilde{C}_5n^{\eta+\alpha-\omega/2}q_n\frac{\log n}{2n}C_n \\
\geq & \min\left\{\log2, \, \frac{\widetilde{C}_4}{\widetilde{C}_3+o(1)}\right\}-2^{-1}\widetilde{C}_5n^{\eta+\alpha-\omega/2}n^{-1}C_nq_n\log n >0
\end{align*}
with probability tending to one as $n\to\infty$, where $\widetilde{C}_5$ is some positive constant, the first inequality follows from the facts that $\log(1+x)\geq \min\{\log2, x/2\}$ for any $x>0$, $N_{\mathcal{S}^{(\ell_{\min})}} =1+q_n\ell_{\min}$, and $\ell_{\min}\leq K_n\leq O(n^{\eta+\alpha-\omega/2})$,  the second inequality uses \eqref{eq: A21}, and the last inequality follows from the assumption that $n^{\eta+\alpha-\omega/2}n^{-1}C_nq_n\log(n)=o(1)$. Therefore, this result, together with the definition of $\hat{\ell}$ in \eqref{eq: ell_hat_def}, entails that $P(\hat{\ell}\geq \ell_{\min})\to 1$ as $n\to\infty$. This completes the proof of Theorem~\ref{Th4}.

\section*{Appendix B: Proofs of Lemmas \ref{lem: gdiff-bound}-\ref{lem: theta-diff-bound}}\label{AppB}

\renewcommand{\theequation}{B.\arabic{equation}}
\setcounter{equation}{0}
\setcounter{section}{0}
\renewcommand{\thesubsection}{B.\arabic{subsection}}

\vspace{4mm}

\noindent{\bf Proof of Lemma~\ref{lem: gdiff-bound}}. We consider $\btheta_{\mathcal{S}}=\btheta_{0\mathcal{S}}+Cq_n^{-1/2}n^{-\alpha}\bu$ with  $\bu\in\mathbb{R}^{|\mathcal{S}|q_n}$ satisfying $\|\bu\|=1$.
Using Knight’s identity given in \citet[page~758]{hknight1998limiting}, we can obtain the following 
result 
\begin{align*}
\rho_{\tau}(u-v)-\rho_{\tau}(u)
=v[I(u<0)-\tau]+\int_0^v[I(u\leq t)-I(u\leq 0)]\,dt.
\end{align*}
Thus, 
we have 
\begin{align*}
& g(\btheta_{\mathcal{S}})-g(\btheta_{0\mathcal{S}})
=E\left[\rho_{\tau}\left(Y-\bz_{\mathcal{S}}^{\top}\btheta_{\mathcal{S}}\right)-\rho_{\tau}(Y)\right]
-E\left[\rho_{\tau}\left(Y-\bz_{\mathcal{S}}^{\top}\btheta_{0\mathcal{S}}\right)-\rho_{\tau}(Y)\right]\nonumber\\
=&E\left[\rho_{\tau}\left(Y-\bz_{\mathcal{S}}^{\top}{\btheta_{\mathcal{S}}}\right)-\rho_{\tau}\left(Y-\bz_{\mathcal{S}}^{\top}{\btheta_{0\mathcal{S}}}\right)\right]\nonumber\\
=&I_1
+E\left\{\int_0^{Cq_n^{-1/2}n^{-\alpha}\bz_{\mathcal{S}}^{\top}\bu}\left[I(Y-\bz_{\mathcal{S}}^{\top}{\btheta_{0\mathcal{S}}}\leq t)-I(Y-\bz_{\mathcal{S}}^{\top}{\btheta_{0\mathcal{S}}}\leq 0)\right]\,dt\right\}\nonumber\\
=&I_1
+E\left\{\int_0^{Cq_n^{-1/2}n^{-\alpha}\bz_{\mathcal{S}}^{\top}\bu}\left[F_{Y|\bx_{\mathcal{S}}}(\bz_{\mathcal{S}}^{\top}{\btheta_{0\mathcal{S}}}+t)-F_{Y|\bx_{\mathcal{S}}}(\bz_{\mathcal{S}}^{\top}{\btheta_{0\mathcal{S}}})\right]\,dt\right\}\nonumber\\
\overset{\Delta}{=}& I_1 + I_2,
\end{align*}
where $I_1=Cq_n^{-1/2}n^{-\alpha} E\left\{\bz_{\mathcal{S}}^{\top}\bu\left[I(Y-\bz_{\mathcal{S}}^{\top}{\btheta_{0\mathcal{S}}}<0)-\tau\right]\right\}$.

By the property of conditional expectation and h\"{o}lder inequality, we have
\begin{align}\label{eq: A2}
|I_1|
=&Cq_n^{-1/2}n^{-\alpha}\left|E\left[E\left\{\bz_{\mathcal{S}}^{\top}\bu\left[I(Y-\bz_{\mathcal{S}}^{\top}{\btheta_{0\mathcal{S}}}<0)-\tau\right]\bigg|\bx_S\right\}\right]\right|\nonumber\\
=& Cq_n^{-1/2}n^{-\alpha}\left|E\left\{\bz_{\mathcal{S}}^{\top}\bu\left[F_{Y|\bx_{S}}(\bz_{\mathcal{S}}^{\top}\btheta_{0\mathcal{S}})-F_{Y|\bx_{S}}(Q_{\tau}(Y|\bx_S))\right]\right\}\right|\nonumber\\
\leq& Cq_n^{-1/2}n^{-\alpha}\left\{E\left(\bz_{\mathcal{S}}^{\top}\bu\right)^2\right\}^{1/2}
\left\{E\left[F_{Y|\bx_{S}}(\bz_{\mathcal{S}}^{\top}\btheta_{0\mathcal{S}})-F_{Y|\bx_{\mathcal{S}}}(Q_{\tau}(Y|\bx_S))\right]^2\right\}^{1/2}.
\end{align}
Write $\bu=(\bu_1^{\top}, \cdots, \bu_{|\mathcal{S}|}^{\top})^{\top}$ with each $\bu_j\in \mathbb{R}^{q_n}$. 
Using the fact of
$\|B_{j\ell}(\cdot)\|_{\infty}\leq 1$ and the Cauchy–Schwarz inequality, we have $\left(\bB_j^T(X_j)\bu_j\right)^2\leq q_n\|\bu_j\|^2$ for each $j=1, \ldots, |\mathcal{S}|$. Using h\"{o}lder inequality entails
\begin{align}\label{eq: A3}
E\left(\bz_{\mathcal{S}}^{\top}\bu\right)^2
=& E\left(\sum_{j\in\mathcal{S}}\bB_j^{\top}(X_j)\bu_j\right)^2
\leq |\mathcal{S}|\sum_{j\in\mathcal{S}}E\left(\bB_j^{\top}(X_j)\bu_j\right)^2 \nonumber\\
\leq & |S|\sum_{j\in\mathcal{S}} E\left(q_n\|\bu_j\|^2\right)
=|S|q_n\|\bu\|^2
=|S|q_n.
\end{align}
It follows from the similar arguments in the proof of Lemma 3.1 in \cite{he2013} that there exist a positive constant $c_4>0$ such that
\begin{align}\label{eq: A4}
E\left[F_{Y|\bx_{\mathcal{S}}}(\bz_{\mathcal{S}}^{\top}\btheta_{0\mathcal{S}})-F_{Y|\bx_{\mathcal{S}}}(Q_{\tau}(Y|\bx_{\mathcal{S}}))\right]^2
\leq c_4|\mathcal{S}|^2q_n^{-2d}.
\end{align}
Combining \eqref{eq: A2}, \eqref{eq: A3}, and \eqref{eq: A4} gives
\begin{align}\label{eq: A5}
|I_1|
\leq Cq_n^{-1/2}n^{-\alpha} (|S|q_n)^{1/2}(c_4 |\mathcal{S}|^2q_n^{-2d})^{1/2}
=O\left(|\mathcal{S}|^{3/2}q_n^{-d}n^{-\alpha}\right).
\end{align}

Next, we consider $I_2$. There exists some $\xi$ between $\bz_{\mathcal{S}}^{\top}\btheta_{0\mathcal{S}}+t$ and $\bz_{\mathcal{S}}^{\top}\btheta_{0\mathcal{S}}$ such that
\begin{align}\label{eq: A6}
I_2
=& E\left[\int_0^{Cq_n^{-1/2}n^{-\alpha}\bz_{\mathcal{S}}^{\top}\bu}f_{Y|\bx_{\mathcal{S}}}(\xi)t\,dt\right]
=O(1)\cdot E\left(Cq_n^{-1/2}n^{-\alpha}\bz_{\mathcal{S}}^{\top}\bu\right)^2 \nonumber\\
=& O(|S|n^{-2\alpha})
=O(r_nn^{-2\alpha}),
\end{align}
where the second equality follows from Condition \ref{con: fy}
and the third equality uses \eqref{eq: A3}. In view of \eqref{eq: A5} and \eqref{eq: A6}, we have $I_1=o(I_2)$ under
the assumption  $r_n^{1/2}q_n^{-d}n^{\alpha}=o(1)$. It follows from Condition \ref{con: fy} that $I_2$ is non-negative. Thus, the conclusion of Lemma~\ref{lem: gdiff-bound} holds.

\vspace{8mm}

\noindent{\bf Proof of Lemma \ref{lem: theta-diff-bound}}.
An direct application of Lemma 2 of \cite{hjort2011asymptotics} yields
\begin{align*}
& P\left(\|\hat{\btheta}_{\mathcal{S}}-\btheta_{0\mathcal{S}}\|\geq Cq_n^{-1/2}n^{-\alpha}\right) \\
\leq & P\left(\sup_{\|\btheta_{\mathcal{S}}-\btheta_{0\mathcal{S}}\|\leq Cq_n^{-1/2}n^{-\alpha}}|g_n(\btheta_{\mathcal{S}})-g(\btheta_{\mathcal{S}})|\geq 2^{-1} \inf_{\|\btheta_{\mathcal{S}}-\btheta_{0\mathcal{S}}\|= Cq_n^{-1/2}n^{-\alpha}}(g(\btheta_{\mathcal{S}})-g(\btheta_{0\mathcal{S}}))\right).
\end{align*}
This together with Lemma \ref{lem: gdiff-bound} entails 
\begin{align}\label{eq: theta-diff}
&P\left(\|\hat{\btheta}_{\mathcal{S}}-\btheta_{0\mathcal{S}}\|\geq Cq_n^{-1/2}n^{-\alpha}\right) \nonumber\\
\leq & P\left(\sup_{\|\btheta_{\mathcal{S}}-\btheta_{0\mathcal{S}}\|\leq Cq_n^{-1/2}n^{-\alpha}}|g_n(\btheta_{\mathcal{S}})-g(\btheta_{\mathcal{S}})|\geq 2^{-1} \inf_{\|\btheta_{\mathcal{S}}-\btheta_{0\mathcal{S}}\|= Cq_n^{-1/2}n^{-\alpha}}(g(\btheta_{\mathcal{S}})-g(\btheta_{0\mathcal{S}}))\right)\nonumber\\
\leq& P\left(\sup_{\|\btheta_{\mathcal{S}}-\btheta_{0\mathcal{S}}\|\leq Cq_n^{-1/2}n^{-\alpha}}|g_n(\btheta_{\mathcal{S}})-g(\btheta_{\mathcal{S}})|\geq C_4r_nn^{-2\alpha}\right)	\nonumber\\
\leq &
P\left(|g_n(\btheta_{0\mathcal{S}})-g(\btheta_{0\mathcal{S}})|\geq 2^{-1}C_4r_nn^{-2\alpha}\right) \nonumber\\
&\quad + P\left(\sup_{\|\btheta_{\mathcal{S}}-\btheta_{0\mathcal{S}}\|\leq Cq_n^{-1/2}n^{-\alpha}}\big|[g_n(\btheta_{\mathcal{S}})-g_n(\btheta_{0\mathcal{S}})]-[g(\btheta_{\mathcal{S}})-g(\btheta_{0\mathcal{S}})]\big|\geq 2^{-1}C_4r_nn^{-2\alpha}\right) \nonumber\\
\overset{\Delta}{=}& J_1 + J_2. 
\end{align}
for all $n$ sufficiently large, where $C_4>0$ is some constant.

We next derive the bounds for the above two probabilities, respectively. We first evaluate $J_1$.
Let $V_i=\rho_{\tau}(Y_i-\bz_{i, \,\mathcal{S}}^{\top}\btheta_{0\mathcal{S}})-\rho_{\tau}(Y_i)$, $i=1, \cdots, n$.  Then $g_n(\btheta_{0\mathcal{S}})-g(\btheta_{0\mathcal{S}})=n^{-1}\sum_{i=1}^n[V_i-E(V_i)]$. 
An application of the inequality \eqref{eq: rho-Lipschitz} entails
$\left|V_i\right|=|\rho_{\tau}(Y_i-\bz_{i, \,\mathcal{S}}^{\top}\btheta_{0\mathcal{S}})-\rho_{\tau}(Y_i)|\leq \left|\bz_{i, \,\mathcal{S}}^{\top}\btheta_{0\mathcal{S}}\right|$. 
It follows from the similar arguments in the proof of Lemma 3.1 in \cite{he2013} that there exist a positive constant $c_5$ such that
\begin{align*}
|Q_{\tau}(Y|\bx_{\mathcal{S}})-\bz_{i, \,\mathcal{S}}^{\top}\btheta_{0\mathcal{S}}|
\leq c_5|\mathcal{S}|q_n^{-d}
\leq c_5r_nq_n^{-d}
=o(r_n^{1/2}n^{-\alpha}).
\end{align*} 
Therefore, it follows from Condition \ref{con: fy} that $|V_i|\leq M_1$ and $\var(V_i)\leq E(V_i^2)\leq M_1^2$ for some positive constant $M_1$.
Applying Bernstein's inequality yields 
\begin{align}\label{eq: J1-bound}
&J_1= P\left(\left|\frac{1}{n}\sum_{i=1}^n[V_i-E(V_i)]\right|\geq 2^{-1}C_4r_nn^{-2\alpha}\right)\nonumber\\
\leq & 2\exp\left\{-\frac{C_4^2r_n^2n^{1-4\alpha}/4}{2(M_1^2+M_12^{-1}C_4r_nn^{-2\alpha}/3)}\right\}
\leq 2\exp\left(-C_5r_n^2n^{1-4\alpha}\right)
\end{align} 
for some positive constant $C_5$.

Next, we will use Massart's concentration theorem \citep{massart2000constants} to evaluate the second term $J_2$. 
Define
\begin{align*}
W_i = \rho_{\tau}\left(Y_i-\bz_{i, \,\mathcal{S}}^{\top}\btheta_{\mathcal{S}}\right)
-\rho_{\tau}\left(Y_i-\bz_{i, \,\mathcal{S}}^{\top}\btheta_{0\mathcal{S}}\right), i=1, \cdots, n.
\end{align*}
Under this definition, the second term $J_2$ is
\begin{align*}
P\left(\sup_{\|\btheta_{\mathcal{S}}-\btheta_{0\mathcal{S}}\|\leq Cq_n^{-1/2}n^{-\alpha}}n^{-1}\left|\sum_{i=1}^n[W_i-E(W_i)]\right|\geq 2^{-1}C_4r_nn^{-2\alpha}\right).
\end{align*}
Using \eqref{eq: rho-Lipschitz} with the fact of
$\|B_{j\ell}(\cdot)\|_{\infty}\leq 1$ and the Cauchy-Schwarz inequality entails
\begin{align}\label{eq: Inequality-Wi}
&\sup_{\|\btheta_{\mathcal{S}}-\btheta_{0\mathcal{S}}\|\leq Cq_n^{-1/2}n^{-\alpha}}|W_i|
\leq  
\sup_{\|\btheta_{\mathcal{S}}-\btheta_{0\mathcal{S}}\|\leq Cq_n^{-1/2}n^{-\alpha}}|\bz_{i, \,\mathcal{S}}^{\top}(\btheta_{\mathcal{S}}-\btheta_{0\mathcal{S}})| \nonumber\\
\leq & \sup_{\|\btheta_{\mathcal{S}}-\btheta_{0\mathcal{S}}\|\leq Cq_n^{-1/2}n^{-\alpha}}\|\btheta_{\mathcal{S}}-\btheta_{0\mathcal{S}}\|_1 
\leq  (r_nq_n)^{1/2} (Cq_n^{-1/2}n^{-\alpha})
=Cr_n^{1/2}n^{-\alpha}
\end{align}
for some positive constant $C$.
Let $\varepsilon_1, \cdots, \varepsilon_n$ be a Rademacher sequence (i.e., independent and identically distributed sequence taking values of $\pm1$ with probability 1/2) independent of
$W_1, \cdots, W_n$. It follows from the symmetrization theorem (Lemma 2.3.1 in \citet{vaart1996weak}) that 
\begin{align}\label{eq: bound1-Wn}
E\left\{\sup_{\|\btheta_{\mathcal{S}}-\btheta_{0\mathcal{S}}\|\leq Cq_n^{-1/2}n^{-\alpha}}n^{-1}\left|\sum_{i=1}^n[W_i-E(W_i)]\right|\right\}
\leq 2E\left[
\sup_{\|\btheta_{\mathcal{S}}-\btheta_{0\mathcal{S}}\|\leq Cq_n^{-1/2}n^{-\alpha}}n^{-1}\left|\sum\limits_{i=1}^n\varepsilon_iW_i\right|\right].
\end{align}
By the contraction theorem \citep{ledoux1991probability} and the Lipschitz property of $\rho_{\tau}(u)$,  
we have  
\begin{align}\label{eq: bound2-Wn}
& E\left[
\sup_{\|\btheta_{\mathcal{S}}-\btheta_{0\mathcal{S}}\|\leq Cq_n^{-1/2}n^{-\alpha}}n^{-1}\left|\sum\limits_{i=1}^n\varepsilon_iW_i\right|\right]
\leq  2E\left[
\sup_{\|\btheta_{\mathcal{S}}-\btheta_{0\mathcal{S}}\|\leq Cq_n^{-1/2}n^{-\alpha}}n^{-1}\left|\sum\limits_{i=1}^n\varepsilon_i\bz_{i, \,\mathcal{S}}^{\top}(\btheta_{\mathcal{S}}-\btheta_{0\mathcal{S}})\right|\right]\nonumber\\
\leq & 2Cq_n^{-1/2}n^{-1-\alpha}E\left[
\left\|\sum\limits_{i=1}^n\varepsilon_i\bz_{i, \,\mathcal{S}}\right\|\right]
\leq 2Cq_n^{-1/2}n^{-1-\alpha}  \left\{E\left[
\left\|\sum\limits_{i=1}^n\varepsilon_i\bz_{i, \,\mathcal{S}}\right\|^2\right]\right\}^{1/2}\nonumber\\
=& 2Cq_n^{-1/2}n^{-1-\alpha}  \left[
\sum\limits_{i_1=1}^n\sum\limits_{i_2=1}^nE(\varepsilon_{i_1}\varepsilon_{i_2}\bz_{i_1, \,\mathcal{S}}^{\top}\bz_{i_2, \,\mathcal{S}})\right]^{1/2}
=2Cq_n^{-1/2}n^{-1-\alpha} \left[
\sum\limits_{i=1}^nE(\varepsilon_{i}^2\bz_{i, \,\mathcal{S}}^{\top}\bz_{i, \,\mathcal{S}})\right]^{1/2}\nonumber\\
=&2Cq_n^{-1/2}n^{-1-\alpha}  \left[
\sum\limits_{i=1}^nE(\bz_{i, \,\mathcal{S}}^{\top}\bz_{i, \,\mathcal{S}})\right]^{1/2}
\leq 2C|\mathcal{S}|^{1/2}n^{-1/2-\alpha}
\leq 2Cr_n^{1/2}n^{-1/2-\alpha},
\end{align}
where the second inequality applies the Cauchy–Schwarz inequality, the third inequality holds from the Jensen's inequality and the last inequality uses the fact of
$\|B_{j\ell}(\cdot)\|_{\infty}\leq 1$.
Combining \eqref{eq: bound1-Wn} and \eqref{eq: bound2-Wn} gives 
\begin{align*}
E\left\{\sup_{\|\btheta_{\mathcal{S}}-\btheta_{0\mathcal{S}}\|\leq Cq_n^{-1/2}n^{-\alpha}}n^{-1}\left|\sum_{i=1}^n[W_i-E(W_i)]\right|\right\}
\leq 4Cr_n^{1/2}n^{-1/2-\alpha}.
\end{align*}
Let $W=\sup\limits_{\|\btheta_{\mathcal{S}}-\btheta_{0\mathcal{S}}\|\leq Cq_n^{-1/2}n^{-\alpha}}n^{-1}\left|\sum_{i=1}^n[W_i-E(W_i)]\right|$. 
By \eqref{eq: Inequality-Wi} and Massart's concentration theorem \citep{massart2000constants},
we have
\begin{align*}
J_2
=&P\left\{W\geq E(W)+ \left[2^{-1}C_4|\mathcal{S}|n^{-2\alpha}-E(W)\right]\right\} \nonumber\\
\leq & P\left\{W\geq E(W)+ \left[2^{-1}C_4r_nn^{-2\alpha}-4Cr_n^{1/2}n^{-1/2-\alpha}\right]\right\}\nonumber\\
\leq& \exp\left\{-\frac{n\left[2^{-1}C_4r_nn^{-2\alpha}-4Cr_n^{1/2}n^{-1/2-\alpha}\right]^2}{2(2Cr_n^{1/2}n^{-\alpha})^2}\right\}
\leq \exp\left(-C_6r_nn^{1-2\alpha}\right) 
\end{align*}
for some positive constant $C_6$.
This, together with \eqref{eq: theta-diff} and \eqref{eq: J1-bound}, completes the proof of Lemma \ref{lem: theta-diff-bound}.

\vspace{2mm}

\section*{Appendix C: Comparison of Computation Time }\label{AppC}

\renewcommand{\theequation}{C.\arabic{equation}}
\setcounter{equation}{0}
\setcounter{section}{0}

\renewcommand{\thesubsection}{C.\arabic{subsection}}

\setcounter{table}{0}
\renewcommand{\thetable}{S\arabic{table}}

Following the suggestion of one referee, for the demonstration purpose, we have
compared the computation time of our method with other approaches in Examples 1 and 2. The corresponding average computation times in seconds over 100 replications are reported in Tables~\ref{tab:Time-Example1} and~\ref{tab:Time-Example2}.

\begin{table}[H]
	\centering
    \caption{Average computation times (in seconds) with standard errors in parentheses for different methods over 100 replications for Example 1 with $(n, p)=(300, 3000)$.} \label{tab:Time-Example1} 
	\begin{tabular}{ccccc}
		\hline
		$\tau$ & QaSIS       & Q-SCAD          & AQFS   & AQFS + QBIC         \\
		\hline
		0.3 & 9.53 (0.29) & 6027.20 (79.99) & 274.49 (0.09) & 275.85 (0.12) \\
		0.5 & 9.93 (0.29) & 6005.73 (62.33) & 275.97 (0.08) & 276.09 (0.13) \\
		0.7 & 9.04 (0.17) & 6087.91 (67.53) & 274.74 (0.14) & 275.55 (0.11) \\
		\hline
	\end{tabular}
\end{table}

\vspace{2mm}
\begin{table}[H]
	\centering
    \caption{Average computation times (in seconds) with standard errors in parentheses for different methods over 100 replications for Example 2 with $(n, p)=(300, 3000)$.} \label{tab:Time-Example2} 
	\begin{tabular}{ccccc}
		\hline
		$\tau$ & QaSIS       & QN-SCAD         & AQFS   & AQFS + QBIC          \\
		\hline
		0.5 & 9.41 (0.23) & 6110.71 (57.49) & 276.67 (0.09) & 276.95 (0.09) \\
		0.7 & 8.79 (0.16) & 6187.21 (75.01) & 275.64 (0.10) & 276.70 (0.08) \\
		\hline
	\end{tabular}
\end{table}

It is observed from Tables S1 and S2 that our screening algorithm AQFS can be more than 22 times faster than the penalized quantile regression methods, although our screening algorithm is slower than the marginal screening method QaSIS.  This is reasonable because 
QaSIS selects relevant covariates at one time while AQFS selects variables sequentially and needs to repeat forward quantile regression $K_n$ times. 
We also observe that the difference between the computation time of AQFS and the computation time of AQFS+QBIC is very small, indicating that the time for the selection stage using QBIC in our method is negligible.


\begin{thebibliography}{99}
	
	\bibitem[Belloni and Chernozhukov, 2011]{belloni2011}
	Belloni, A. and Chernozhukov, V. (2011). $\ell_1$-penalized quantile regression in high-dimensional sparse models. {\it The Annals of Statistics}, {\bf 39}, 82--130.

	
	
	\bibitem[Cheng et al., 2018]{cheng2018greedy}
	Cheng, M.-Y., Feng, S., Li, G., and Lian, H. (2018). 
	Greedy forward regression for variable screening.
	{\it Australian $\&$ New Zealand Journal of Statistics}, {\bf 60}, 20--42.  
	
	
	\bibitem[Cheng et al., 2016]{cheng2016forward}
	Cheng, M.-Y., Honda, T., and Zhang, J.-T. (2016). 
	Forward variable selection for sparse ultra-high dimensional varying coefficient models.
	{\it Journal of the American Statistical Association}, {\bf 111}, 1209--1221. 


	\bibitem[Cheng et al., 2011]{cheng2011efficient}
    Cheng, Y., De Gooijer, J. G, and Zerom, D. (2011). 
    Efficient estimation of an additive quantile regression model.
    {\it Scandinavian Journal of Statistics}, {\bf 38}, 46--62. 	
	
	
	\bibitem[De Gooijer and Zerom, 2003]{de2003additive}
	De Gooijer, J. G. and Zerom, D. (2003). 
	On additive conditional quantiles with high-dimensional covariates.
	{\it Journal of the American Statistical Association}, {\bf 98}, 135--146. 


	
	\bibitem[Fan et al., 2011]{fan2011}
	Fan, J., Feng, Y., and Song, R. (2011). 
	Nonparametric independence screening in sparse ultra-high dimensional additive models. {\it Journal of the American Statistical Association}, {\bf 106}, 544--557.
	
	
	\bibitem[Fan and Li, 2001]{fan2001variable}
	Fan, J. and Li, R. (2001). Variable selection via nonconcave penalized likelihood and its oracle properties. {\it Journal of the American statistical Association}, {\bf 96}, 1348--1360.	
	
	\bibitem[Fan and Lv, 2008]{fanlv2008}
	Fan, J. and Lv, J. (2008). Sure independence screening for ultrahigh dimensional feature space. {\it Journal of the Royal Statistical Society, Series B}, {\bf 70}, 849--911.
	
	\bibitem[Fan and Lv, 2010]{fan2010selective}
	Fan, J. and Lv, J. (2010). A selective overview of variable selection in high dimensional feature space. {\it Statistica Sinica}, {\bf 20}, 101--148.
	
	
	
	\bibitem[Fan et al., 2009]{fan2009ultrahigh}
	Fan, J., Richard, S., and Wu, Y. (2009). 
	Ultrahigh dimensional feature selection: beyond the linear model. {\it The Journal of Machine Learning Research}, {\bf 10}, 2013--2038.
	
	
	
	
	
	\bibitem[Fenske et al., 2011]{Fenskeetal2011GAMBOOST}
	Fenske, N., Kneib, T. and Hothorn, T. (2011). Identifying risk factors 
	for severe childhood malnutrition by boosting additive quantile 
	regression. {\it Journal of the American Statistical Association}, {\bf 106}, 494--510.
	
	
	\bibitem[He et al., 2013]{he2013}
	He, X., Wang, L., and Hong, H. G. (2013). Quantile-adaptive model-free variable screening for high-dimensional heterogeneous data. {\it The Annals of Statistics}, {\bf 41}, 342--369.
	
	\bibitem[Hjort and Pollard, 2011]{hjort2011asymptotics}
	Hjort, N. L. and Pollard, D. (1993). Asymptotics for minimisers of convex processes. Technical report, Department of Statistics, Yale University.
	Available at http://www.stat.yale.edu/~pollard/Papers/convex.pdf.
	
	
	\bibitem[Hoeffding, 1963]{hoeffding1963probability}
	Hoeffding, W. (1963). 
	Probability Inequalities for Sums of Bounded Random Variables.
	{\it Journal of the American Statistical Association}, {\bf 58}, 13--30.
	
	
	
	
	
	\bibitem[Horowitz and Lee, 2005]{Horowitz2005nonparametric}
	Horowitz, J. and Lee, S. (2005). 
	Nonparametric Estimation of an Additive Quantile Regression Model.
	{\it Journal of the American Statistical Association}, {\bf 100}, 1238--1249.
	
	
	
	\bibitem[Huang et al., 2010]{huang2010variable}
	Huang, J., Horowitz, J., and Wei, F. (2010). 
	Variable selection in nonparametric additive models.
	{\it The Annals of Statistics}, {\bf 38}, 2282--2313.
	
    \bibitem[Jiang at al., 2020]{jiang2020functional}
    Jiang, F., Cheng, Q., Yin, G., and Shen, H. (2020). Functional Censored Quantile Regression.
   {\em Journal of the American Statistical Association}~{\bf 115}, 931--944.	
	
	\bibitem[Knight, 1998]{hknight1998limiting}
	Knight, Keith. (1998). Limiting distributions for $L_1$ regression estimators under general conditions. 
	{\it The Annals of Statistics}, {\bf 26}, 755--770.
	

	
	\bibitem[Kong et al., 2017]{kong2017}
	Kong, Y., Li, D., Fan, Y. and Lv, J. (2017). Interaction pursuit in high-dimensional multi-response regression via distance correlation. {\it The Annals of Statistics}, {\bf 45}, 897--922.
	

	
	
	\bibitem[Lee et al., 2014]{LeeNohPark2014}
	Lee, E.R., Noh, H. and Park, B.U. (2014). Model selection via Bayesian information criterion for quantile regression models. {\it Journal of the American Statistical Association}, {\bf 109}, 216--229.
	
	\bibitem[Li et al., 2022]{Li2022high}
	Li, D., Kong, Y., Fan, Y., and Lv, J. (2022). High-dimensional interaction detection with false sign rate control. {\it Journal of Business $\&$ Economic Statistics}, {\bf 40}, 1234--1245.
	
	

	
	
	\bibitem[Li et al., 2012]{li2012feature}
	Li, R., Zhong, W. and Zhu, L. (2012). Feature screening via distance correlation learning. {\it Journal of the American Statistical Association}, {\bf 107}, 1129-1139.	
	
	
	
	\bibitem[Li et al., 2017]{li2017profile}
	Li, Y., Li, G., Lian, H., and Tong, T. (2017). 
	Profile forward regression screening for ultra-high dimensional semiparametric varying coefficient partially linear models.
	{\it Journal of Multivariate Analysis}, {\bf 155}, 133--150.
	

	
	\bibitem[Ledoux and Talagrand, 1991]{ledoux1991probability}
	Ledoux, Michel and Talagrand, Michel (1991). {\it Probability in Banach Spaces: Isoperimetry and Processes}, Springer, Berlin.
	

	
	\bibitem[Ma et al., 2017]{maetal2017}
	Ma, S., Li, R. and Tsai, C. L. (2017). Variable screening via quantile partial correlation. {\it Journal of the American Statistical Association}, {\bf 112}, 650--663.
	

	
	\bibitem[Massart, 2000]{massart2000constants}
	Massart, P. (2000). About the constants in Talagrand's concentration inequalities for empirical processe. {\it The Annals of Probability}, {\bf 28}, 863--884.
	

	
	\bibitem[Serfling, 1980]{serfling1980approximation}
	Serfling, R. (1980). {\it Approximation Theorems of Mathematical Statistics}. Wiley, New York.
	
	\bibitem[Ni and Fang, 2016]{ni2016entropy}
	Ni, L. and Fang, F. (2016). Entropy-based model-free feature screening for ultrahigh-dimensional multiclass classification. {\it Journal of Nonparametric Statistics}, {\bf 28}, 515--530.


	
	\bibitem[Sherwood and Maidman, 2022]{sherwood2022}	
	Sherwood, B. and Maidman, A. (2022). Additive nonlinear quantile regression in ultra-high
	dimension. {\it Journal of Machine Learning Research}, {\bf 23}, 1--47.
	
	
	\bibitem[Sherwood and Wang, 2016]{sherwood2016}
	Sherwood, B. and Wang, L. (2016). Partially linear additive quantile regression in ultra-high dimension. {\it The Annals of Statistics}, {\bf 44}, 288--317.
	

	\bibitem[Song et al., 2014]{song2014censored}
    Song, R., Lu, W., Ma, S., and Jeng, J. X. (2016). Censored rank independence screening for high-dimensional survival data. {\it Biometrika}, {\bf 101}, 799--814.
		
	
	\bibitem[Stone, 1985]{stone1985additive}
	Stone, C. (1985). Additive regression and other nonparametric models. {\it The Annals of Statistics}, {\bf 13}, 689--705.	
	
	\bibitem[Tibshirani, 1996]{tibshirani1996regression}
	Tibshirani, R. (1996). Regression shrinkage and selection via the lasso. {\it Journal of the Royal Statistical Society, Series B}, {\bf 58}, 267--288.	
	
	\bibitem[Vaart and Wellner, 1996]{vaart1996weak}
	Vaart, Aad W and Wellner, Jon A. (1996). {\it Weak Convergence and Empirical Processes: with Applications to Statistics}. Springer, New York.
	
	\bibitem[Votavova et al., 2011]{votavova2011}
	Votavova, H., Dostalova Merkerova, M., Fejglova, K., Krejcik, Z., 
	Pastorkova, A., Tabashidze, N., Topinka, J., Veleminsy, M., JR., Sram, R. J. and 
	Brdicka, R. (2011). Transcriptome alterations in maternal and fetal cells induced by tobacco smoke. {\it Placenta}, {\bf 32}, 763--770.
	
	\bibitem[Wang, 2009]{wang2009}
	Wang, H. (2009). Forward regression for ultra-high dimensional variable screening. {\it Journal of the American Statistical Association}, {\bf 104}, 1512--1524.
	
	\bibitem[Wang et al., 2012]{wangwuli2012}
	Wang, L., Wu, Y. and Li, R. (2012). Quantile regression for analyzing heterogeneity in ultra-high dimension. {\it Journal of the American Statistical Association}, {\bf 107}, 214--222.
	
	
	\bibitem[Wu and Yin, 2015]{wuyin2015}
	Wu, Y. and Yin, G. (2015). Conditional quantile screening in ultrahigh-dimensional heterogeneous data. {\it Biometrika}, {\bf 102}, 65--76.


	
	\bibitem[Zhang, 2010]{zhang2010nearly}
	Zhang, C.H. (2010). Nearly unbiased variable selection under minimax concave penalty. {\it The Annals of Statistics}, {\bf 38}, 894--942.
	

	\bibitem[Zhong et al., 2020]{zhong2020forward}
	Zhong, W., Duan, S., and Zhu, L. (2020). 
	Forward additive regression for ultrahigh dimensional nonparametric additive models.
	{\it Statistica Sinica}, {\bf 30}, 175--192.	
	
	\bibitem[Zhong et al., 2023]{zhong2023feature}
	Zhong, W., Qian, C., Liu, W., Zhu, L., and Li, R. (2023). 
	Feature Screening for Interval-Valued Response with Application to Study Association between Posted Salary and Required Skills.
	{\it Journal of the American Statistical Association}, {\bf 118}, 805--817.	
			
	
	
\end{thebibliography}
\end{document}